\documentclass[reprint, twocolumn,amsmath,amssymb,showpacs,pra,superscriptaddress,aps]{revtex4-1}

\usepackage{epsfig}
\usepackage{color}
\usepackage{array}
\usepackage{graphicx}
\usepackage{bm}
\usepackage{epstopdf}

\DeclareMathOperator\erfc{erfc}

\begin{document}

\title{Engineering momentum profiles of cold-atom beams}

\author{D~Hudson \surname{Smith}}
\affiliation{Clemson University, Clemson, South Carolina 29634, USA}

\author{Artem~G \surname{Volosniev}}
\affiliation{Institut f{\"u}r Kernphysik, Technische Universit{\"a}t Darmstadt, 64289 Darmstadt, Germany}

\date{\today}

\begin{abstract}
We describe a procedure for engineering beams of cold atoms by selectively draining particles from a trapped gas based on momenta. Atoms escape through a filter potential that only transmits atoms with the desired momenta. We outline an algorithm that outputs a filter potential that produces a pre-specified beam momentum profile. We illustrate this procedure for the case of a narrow band-pass (NBP) quantum filter. Lastly, we discuss the application of the NBP filter for probing the self-energy and effective mass of Bose polarons, as well as the corresponding Landau criterion.
\end{abstract}

\maketitle

{\it Introduction. --}   
In this article we discuss a procedure for creating cold-atom beams with momentum transport profiles that can be selected for the matter at hand.
Such beams would enable novel scattering experiments with quantum gases. In particular, 
they could be used to measure parameters that define
few- and many-body physics of cold-atoms systems, e.g., the
scattering length, three-body parameter~\cite{braaten2006, bloch2008} or
the self-energy and the effective mass of a polaron~\cite{massignan2014, schmidt2018}.
The beams can also be used as the initial state for atom interferometry~\cite{Impens2006,Cronin2009}.

\begin{figure}
\centerline{\includegraphics[scale=0.4]{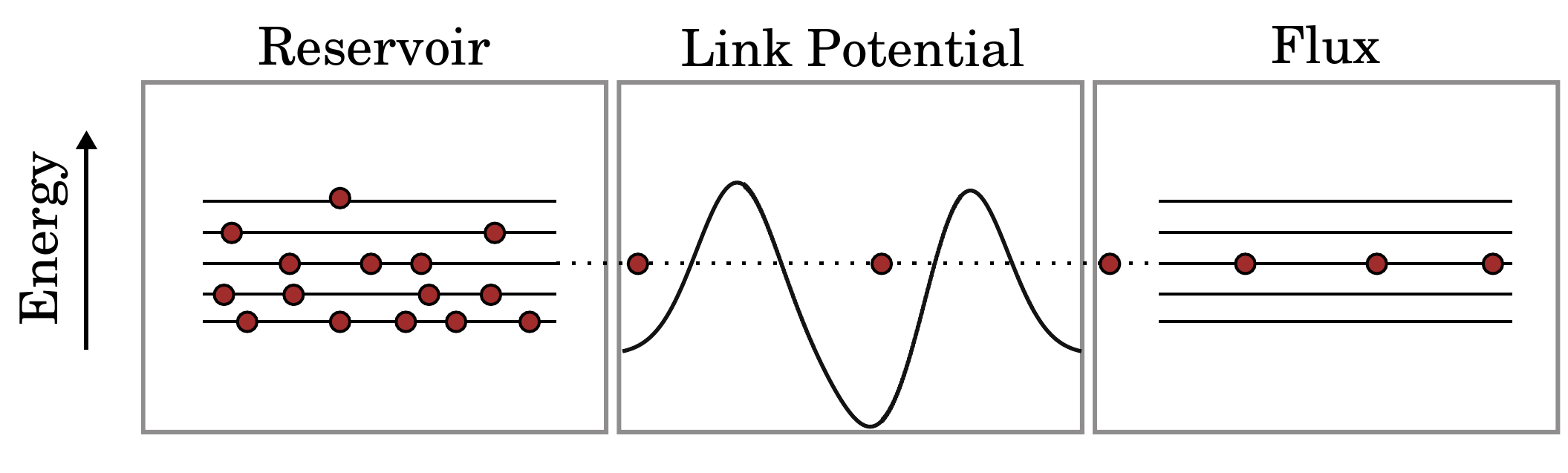}}
\caption{An illustration of the proposal: A reservoir that contains particles of various momenta is connected to an external link potential. The potential filters out the desired momenta, and the particles in the flux region have a known momentum distribution -- here, the distribution is non-zero only in the neighborhood of a chosen momentum. The link potential is a narrow band-pass filter.
}
\label{fig:Figure1}
\end{figure}

Figure ~\ref{fig:Figure1} summarizes our proposal. Analogous to a quantum switch device (`transistor')~\cite{zoller2004, marchukov2016,thuberg2017}, the flux of particles from the `source' (reservoir) is determined by the `gate' (link potential). However, rather than simply controlling the overall transmission rate, our proposal allows one to design the momentum profile of the outgoing flux.
 For simplicity, we illustrate our idea using a one-dimensional geometry (1D) (though our formalism also applies directly to a cylindrically symmetric three-dimensional geometry). We assume that the particles in the reservoir are non-interacting~\footnote{Strong interactions can lead to the transmission behavior that is not captured by the one-body Schr{\"o}dinger equation, e.g.,to a collective resonant transport~\cite{Schlagheck2005}.}, so that their scattering properties may be calculated using the one-body Schr{\"o}dinger equation:
\begin{equation}
-\frac{\hbar^2}{2m}\frac{\partial^2}{\partial x^2}\psi+V_0(x)\psi=\frac{\hbar^2k^2}{2m}\psi,
\label{eq:schr}
\end{equation}
where $m$ is the mass of a particle from the reservoir, $\frac{\hbar^2k^2}{2m}$ is its energy. The link potential $V_0(x)$ produces the transmission coefficient, $T_0(k)$. By carefully tuning $V_0(x)$ one produces a $T_0(k)$ that allows only particles with desired momenta to tunnel through the barrier into the flux region as required by the experimental application. The tunability of $V_0(x)$ in cold-atom set-ups has been recently demonstrated~\cite{esslinger2015,esslinger2018,esslinger2019}, 
suggesting that our proposal relies only on the toolbox available in current cold-atom laboratories.
Note that the momentum profile of the outgoing beam can be measured using single-atom momentum resolution techniques (e.g.,~\cite{jochim2018}), allowing one to confirm that the beam has the desired flux profile.

In this article, we discuss how to determine an appropriate $V_0(x)$ for a given desired flux profile $T_0(k)$. Furthermore, we briefly discuss the 1D Bose polaron problem as a possible application of cold-atom beams. 

Other studies related to the engineering of atomic beams have focused on  the  outcoupling  of  atoms  from a Bose-Einstein condensate \cite{PhysRevLett.97.200402, PhysRevA.80.041605, buning2010slow} or on controlling the transmission of atoms with careful manipulation of an optical lattice \cite{PhysRevLett.84.399, PhysRevLett.107.230401, PhysRevA.92.033614}. In contrast, the current article proposes a method for engineering atomic beams by coupling a trapped gas of atoms to a combination of individually tuned lasers. Moreover, the proposed scheme is generic in the sense that one can approximately produce arbitrary desired momentum profiles of the transmitted atom beam.

%%%%%%%%%%%%%%%%%%%%%%%%%%%%%%%%%%%%%%%%%%%%%%%%%%%%%%%%%%%%%%%%%%%%%%%%%%%%%%%%%%%%%%%%%%%%%%%%%%%%%%%%%%%%%%%%%%%%%%%
%%%%%%%%%%%%%%%%%%%%%%%%%%%%%%%%%%%%%%%%%%%%%%%%%%%%%%%%%%%%%%%%%%%%%%%%%%%%%%%%%%%%%%%%%%%%%%%%%%%%%%%%%%%%%%%%%%%%%%%

\begin{figure}
   \includegraphics[width=1\linewidth]{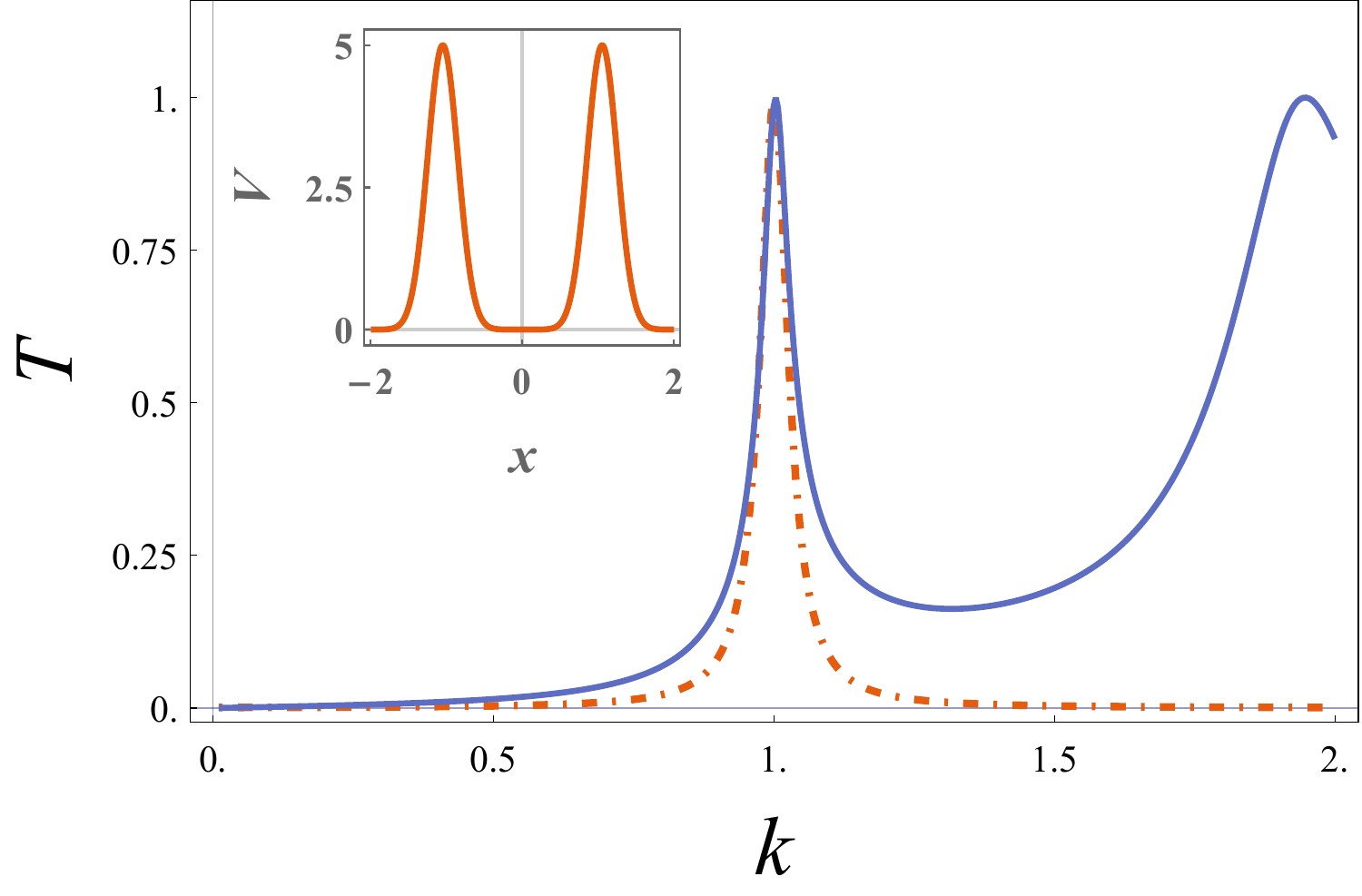}
 \caption[Narrow band-pass filter link potential]{A two-Gaussian solution for the narrow band-pass filter transport profile. The main figure shows the target profile (dot-dashed, red) used during optimization (see Eq.~(\ref{eq:Ttarget})) and the actual transport profile resulting from the optimization procedure (solid, blue). The inset figure shows the optimal link potential $V(x, \bm{\theta^*})$.}
 \label{fig:method_illustration}
\end{figure}

{\it Procedure for Finding a Link Potential. --} 
We find an appropriate link potential $V_0(x) \equiv V(x,\bm{\theta}^*)$ by performing a global search over a family of possible potentials $V(x,\bm{\theta})$ for the parameters $\bm{\theta}^*$ that reduce the $k$-integrated squared error between the desired transmission-momentum profile $T_0(k)$ and the actual profile $T_{\bm\theta}(k)$ produced by a sample potential $V(x, \bm{\theta})$. Concretely, we minimize the cost
\begin{equation}\label{eq:cost1}
  J_{\bm{\theta}} = \sum_{0<k<k_F} w_k\left|T_0(k) - T_{\bm{\theta}}(k)\right|^2,
\end{equation}
where the $k$-integral has been approximated (up to a constant factor) by a sum over a discrete set of momentum values, the interval of integration is $[0,k_F]$, and $w_k$ is the weight given to momentum $k$. The weights are chosen to emphasize or de-emphasize special regions of $k$ during the minimization. For instance, for a narrow band-pass filter that forbids transmission for all $k$ except in the neighborhood of a chosen value $k_0$ (see Figs.~\ref{fig:Figure1} and~\ref{fig:method_illustration}), it is appropriate to increase the weights in the region of $k_0$. The convergence of our approach is somewhat sensitive to the choice of $w_k$.

We developed a strategy for choosing the weights that takes into account the following considerations: {\it i)} the transport coefficient is zero for $k=0$ 
(there can be no transmission for $k=0$ and $V\neq 0$ in 1D), so the weight can be smaller for small $k$, {\it ii)} the transport coefficient approaches 1 for large $k$ so higher weights are required in the large-$k$ region if one wishes to suppress flux at large $k$, and {\it iii)} to reproduce narrow features in the target transport profile, it may be helpful to increase the weight in the $k$-region of these features. Following these principles, we arrive at the following formula for the weight function:
\begin{equation}\label{eq:weight-of-k}
  w(k;r) = w_{\mathrm{bg}}(k) + rT_0(k)
\end{equation}
where $w_{\mathrm{bg}}(k)$ is chosen to account for considerations {\it i)} and {\it ii)} above, and the term proportional to the positive constant $r$ accounts for consideration {\it iii)}. The form  of $w_{\mathrm{bg}}(k)$ can be inferred from typical transmission coefficients. For convenience, we use the analytic form for the transmission coefficient produced by
the Morse potential $\hbar^2k_0^2/[2m \cosh^2(k_0x/\sqrt{2})]$ (cf.~\cite{landau1977}):
\begin{equation}\label{eq:weight-of-k}
  w_{\mathrm{bg}}(k) = \left[\frac{\sinh^2(\sqrt{2}\pi k/k_0)}{\sinh^2(\sqrt{2}\pi k/k_0) + \cosh^2(\sqrt{7}\pi/2)}\right]^{1/2},
\end{equation}
where the parameter $k_0$ determines a typical energy scale (for an example see our illustration below).
The parameter $r$ was chosen by trial and error for each target profile $T_0(k)$.

With the goal of discovering experimentally viable solutions, we parameterize the family of link potentials $V(x, \bm{\theta})$, as a sum of $N$ Gaussians, each of the form
\begin{equation}\label{eq:V-param}
V_i(x; A_i, \mu_i, \sigma_i) = \frac{A_i}{\sqrt{2\pi\sigma_i^2}}\exp\left[{-\frac{(x-\mu_i)^2}{2\sigma_i^2}}\right];
\end{equation}
the parameter $\bm{\theta}$ denotes the parameter space $\{A_1, \mu_1, \sigma_1,...,A_N, \mu_N, \sigma_N\}$. While minimizing Eq.~\eqref{eq:cost1}, we enforce the parameter constraints listed in Table \ref{tab:constraints}. In addition to these constraints on the parameters, we enforce a constraint requiring that the link potential should not extend beyond the region of potential support $x\in[-x_0,x_0]$. 
To accomplish this, we minimize the boundary-augmented cost function
\begin{equation}\label{eq:cost2}
  J_{\bm{\theta}}^{\mathrm{aug}} = J_{\bm{\theta}} + \alpha \sum_{i=1}^N\int\limits_{|x|>x_0}dx\,|V_i(x; A_i,\mu_i,\sigma_i)|^2,
\end{equation}
where $\alpha$ is a tuning parameter chosen empirically to aid in convergence. 
 If the potential $V_i$ is outside $[-x_0,x_0]$ then the boundary-augmented 
cost function evaluates to $J_{\bm{\theta}}\simeq J_{\bm{\theta}} +\alpha A_i$, which means that $\alpha$
should be chosen of the order $J_{\bm{\theta}}/A_{\mathrm{min}}$ to dictate the constraint.
 The integral in Eq.~\eqref{eq:cost2} evaluates to the complementary error function. The full form of $J_{\bm{\theta}}^{\mathrm{aug}}$ is given in Appendix \ref{app:J}.

\begin{table}[t]
  \renewcommand*{\arraystretch}{1.4}
  \begin{tabular}{m{3cm}|m{5.5cm}}
    Constraints & Experimental Rationale \\
    \hline\hline
    $\sum_{i=1}^{N}\mu_i = 0$ & The cost function has a continuous degeneracy associated with overall translations of the link potential. \\
    \hline
    $\sigma_{\mathrm{min}} \leq \sigma_j \leq \sigma_{\mathrm{max}} $ & Laser beam widths fall between a minimum and maximum value.\\
    \hline
    $A_{\mathrm{min}} \leq |A_j| \leq A_{\mathrm{max}}$ & Laser amplitudes fall between a minimum and maximum value.
  \end{tabular}
  \caption{The explicit constraints on the potential parameters and the rationale for each constraint. The values of $\sigma_{\mathrm{min}}$, $\sigma_{\mathrm{max}}$, $A_{\mathrm{min}}$, and $A_{\mathrm{max}}$ must be determined from the experimental context.}
  \label{tab:constraints}
\end{table}

For a particular choice of $\bm{\theta}$ (and hence $V(x;\bm{\theta})$), we solve for $T_{\bm{\theta}}(k)$ by integrating the Schr{\"o}dinger equation~(\ref{eq:schr}) across the region of the potential and calculating the ratio of the transmitted to the incident flux. In order to do this efficiently, we discretize the second derivative in Eq.~(\ref{eq:schr}), which transforms Eq.~(\ref{eq:schr}) into
a banded linear system of equations solvable in $O(M)$ time where $M$ is the number of $x$-steps. Using these techniques we are able to evaluate the transmission coefficient 3.7 thousand times per second on a 7th generation Intel Core i7 processor for a test involving 1,000 randomly generated two-Gaussian potentials and 100 different scattering momenta. 

%%%%%%%%%%%%%%%%%%%%%%%%%%%%%%%%%%%%%%%%%%%%%%%%%%%%%%%%%%%%%%%%%%%%%%%%%%%%%%%%%%%%%%%%%%%%%%%%%%%%%%%%%%%%%%%%%%%%%%% 

%%%%%%%%%%%%%%%%%%%%%%%%%%%%%%%%%%%%%%%%%%%%%%%%%%%%%%%%%%%%%%%%%%%%%%%%%%%%%%%%%%%%%%%%%%%%%%%%%%%%%%%%%%%%%%%%%%%%%%%

We minimize $J_{\bm{\theta}}^{\mathrm{aug}}$ for $\bm{\theta}^*$ using the global optimization routine called Differential Evolution (DE) \cite{storn1997differential}. This evolutionary-based search algorithm is suitable given the non-convex (multiple local minima) nature of the optimization problem and the continuity of the parameter space. Despite its simplicity, DE does a good job of balancing exploration of the space of link potentials against the need to efficiently learn from each sample with little tuning of the model settings. Empirically, we found DE to converge to good solutions much more quickly than random search perhaps due to DE's ability to incorporate information from the previous iteration of the optimization algorithm. 

{\it Narrow band-pass (NBP) filter transport profile. --}
To illustrate the method described above we optimize for a NBP filter transport profile sharply-peaked near $k=k_0$. 
For our target transport profile, we use a Lorentz profile~(see Figure~\ref{fig:method_illustration})
\begin{equation}\label{eq:Ttarget}
  T(k; k_0,b) = \left[1 + \frac{(k-k_0)^2}{b^2}\right]^{-1}
\end{equation}
where $k_0$ determines the peak position and $b$ determines the width. 
For simplicity, in this subsection we adopt the units $k_0=\hbar=2m=1$, which scales $k_0$
out of the problem. The value of $b$ must be much smaller than $k_0$ 
to have a well-defined peak, but not too small to have realistic time scales
for a one-body tunneling. We set $b=0.03 k_0$, which for a reasonable assumption 
$\hbar^2k_0^2/(2m)=k_B\times \mu$K, where $k_B$ is the Boltzmann constant leads to 
the time scale associated with the resonance width $\frac{2m}{\hbar b^2}\sim 10$ms.

 For the constraints shown in Table~\ref{tab:constraints}, we use $\sigma_{\mathrm{min}}=0.2$, $\sigma_{\mathrm{max}}=3$, $A_{\mathrm{min}}=5$, and $A_{\mathrm{max}}=30$. We set $k_F=2$. We further simplify the optimization by searching over two-Gaussian link-potentials with equal amplitudes and widths. We anticipate that this potential might be the easiest to realize in the laboratory. 
Moreover, it allows us to give a physical interpretation in terms of quasi-discrete energy levels supported by the link.  Even though we work here with a very simple example with only three unknown parameters, a method for globally searching the space of possible link potentials is still required, because the cost function for the optimization has many local minima corresponding to the many ways to produce a resonance state near the scattering energy~$k_0^2$. Moreover this global search technique extends to more complicated transport profiles which necessitate more complicated families of link potentials.

Figure \ref{fig:method_illustration} shows the link potential and transport profile resulting for the NBP filter optimization. The solution suppresses transport except near $k=1$ as set by our Lorentz target profile and near $k=2$ resulting from a second resonance in the scattering potential. 
In general, the transmission coefficient will always be non-zero to the right of the target profile. 
This, however, need not be a problem if the atoms are sourced from a thermal reservoir with sufficiently low population to the right of the target profile~\footnote{For a reservoir at a finite temperature it is beneficial to include the energy distribution, $n(k)$,
in the cost function. To this end, one should simply convolute $T_{\bm{\theta}}(k)$ in Eq.~(\ref{eq:cost1}) with $n(k)$. 
The function $n(k)$ is determined by the Bose-Einstein or Fermi-Dirac distributions, temperature and the shape of the trap. For simplicity, 
we assume in current Eq.~(\ref{eq:cost1}) a one-dimensional Fermi gas at zero temperature in a harmonic trap, i.e., $n(k<k_F)=1$ and zero otherwise.}.
Moreover, the transmission coefficient can be further shaped by using a second NBP filter.
 
\begin{figure}
   \includegraphics[width=1\linewidth]{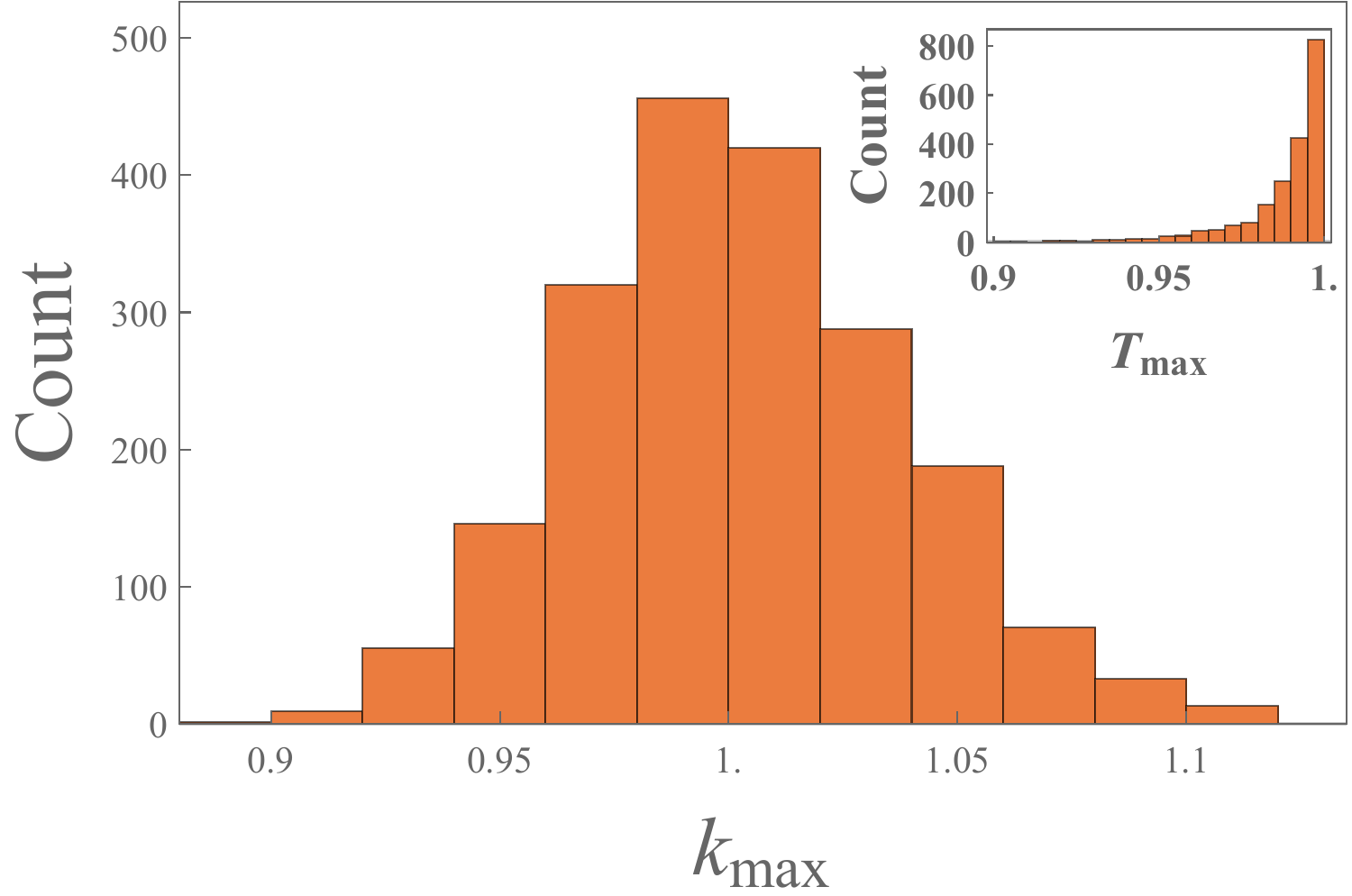}
 \caption[Sensitivity Plot]{Sensitivity of the transport profile to perturbations of the potential parameters. Shown are the distributions of the peak position $k_{\mathrm{max}}$ (main graph) and peak height $T_{\mathrm{max}}$ (inset) for 2000 potentials with parameters randomly perturbed on the order of $5\%$ from the optimized potential presented in Fig.~\ref{fig:method_illustration}.}
 \label{fig:sensitivity}
\end{figure}

It may be experimentally problematic if the transport profile shown in Fig.~\ref{fig:method_illustration} were highly sensitive to the potential parameters. Such sensitivity would require extremely fine control over the laser amplitudes, positions, and widths in order to produce the desired transport profile. To test this sensitivity, we generate 2000 perturbed potentials by varying the six parameters of the two-Gaussian solution shown in Fig.~\ref{fig:method_illustration} by a random-normal multiplicative factor with mean 1 and standard deviation 0.05. The distributions of the peak positions and heights for the 2000 perturbed potentials are shown in Fig.~\ref{fig:sensitivity}. Both the peak positions, $k_{\mathrm{max}}$, and the peak heights, $T_{\mathrm{max}}$, undergo perturbations on the scale of the $5\%$  potential perturbations suggesting that transport properties are relatively insensitive to slight errors in the potential parameters. 

Though we have demonstrated our optimization method in a very simple scenario, it is possible to apply this technique to more complicated scenarios such as a double band-pass filter or step transport profiles. These more complicated transport profiles require more than two Gaussian potentials,
because they rely on multipath interference that must suppress tunneling for certain values of momenta. In our explorations, we found that link potentials made of 3- or 4-Gaussian potentials tended to be more sensitive to random variations of potential parameters. If such potentials are needed to produce the desired transport profile, it may be possible to further augment the cost function in order to preference solutions that are less sensitive to potential perturbations. We leave the thorough exploration of these ideas to future work. In the next section, we discuss a possible experimental application of the NBP filter.

%%%%%%%%%%%%%%%%%%%%%%%%%%%%%%%%%%%%%%%%%%%%%%%%%%%%%%%%%%%%%%%%%%%%%%%%%%%%%%%%%%%%%%%%%%%%%%%%%%%%%%%%%%%%%%%%%%%%%%% 

%%%%%%%%%%%%%%%%%%%%%%%%%%%%%%%%%%%%%%%%%%%%%%%%%%%%%%%%%%%%%%%%%%%%%%%%%%%%%%%%%%%%%%%%%%%%%%%%%%%%%%%%%%%%%%%%%%%%%%%

{\it Application. --} A Bose or Fermi gas can be placed in the flux region  (see Fig.~\ref{fig:Figure1}) to study quantum environments with neutral, mobile impurities -- an important research venue promoted by cold-atom simulators~\cite{zwierlein2009,salomon2009,grimm2012, widera2012, catani2012, fukuhara2013, hu2016,arlt2016,zaccanti2017,ardila2018}. With our proposal, it is possible to investigate the dynamics of impurities that initially have a known momentum profile, thus allowing for a direct measurement of the effective mass and the critical momentum. A detailed discussion of these concepts is beyond the scope of this article. Still, we find it useful to briefly explain them in connection to our proposal. To this end, we consider a degenerate one-dimensional Bose gas with an impurity of momentum $P$. To model this system, we employ a non-linear Schr{\"o}dinger equation for a Bose gas with an impurity atom (See Appendix \ref{app:impurity_bose_gas}). This equation was solved analytically in the context of a nonlinear flow past an obstacle~\cite{hakim1997}, which allows us to work out all properties of the dressed particle in a simple manner; note that this (or a similar) non-linear equation was discussed in Refs.~\cite{kamenev2016, volosniev2017, mistakidis2018, dehkharghani2018, pastukhov2018,pastukhov2019}; see, also,~\cite{sacha2006, catani2012, kain2016, parisi2017,grusdt2017, pastukhov2017, kain2018} for other relevant studies. 

The lowest energy state of the non-linear Schr{\"o}dinger equation with a given $P$ is a combination of two solitons. They make a dissipationless defect in the Bose gas, which accompanies the impurity. The corresponding lowest energy is given by $E\simeq E_B+\epsilon+P^2/(2m_{\mathrm{eff}})$, where $E_B$ is the energy of the gas without an impurity, $\epsilon$ is the self-energy of the dressed particle, and $m_{\mathrm{eff}}$ is its effective mass. The solution is stable only for $P<P_c$; impurities with $P>P_c$ generate grey solitons (cf.~\cite{hakim1997}). Note that quantum fluctuations lead to a finite dissipation (cf.~\cite{astrakharchik2004,sykes2009,Cherny2012}) even for $P<P_c$. We do not consider this effect as it does not change our qualitative presentation.

%%%%%%%%%%%%%%%%%%%%%%%%%%%%%%%%%%%%%%%%%%%%%%%%%%%%%%%%%%%%%%%%%%%%%%%%%%%%%%%%%%%%%%%%%%%%%%%%%%%%%%%%%%%%%%%%%%%%%%%
\begin{figure}
\centerline{\includegraphics[scale=0.3]{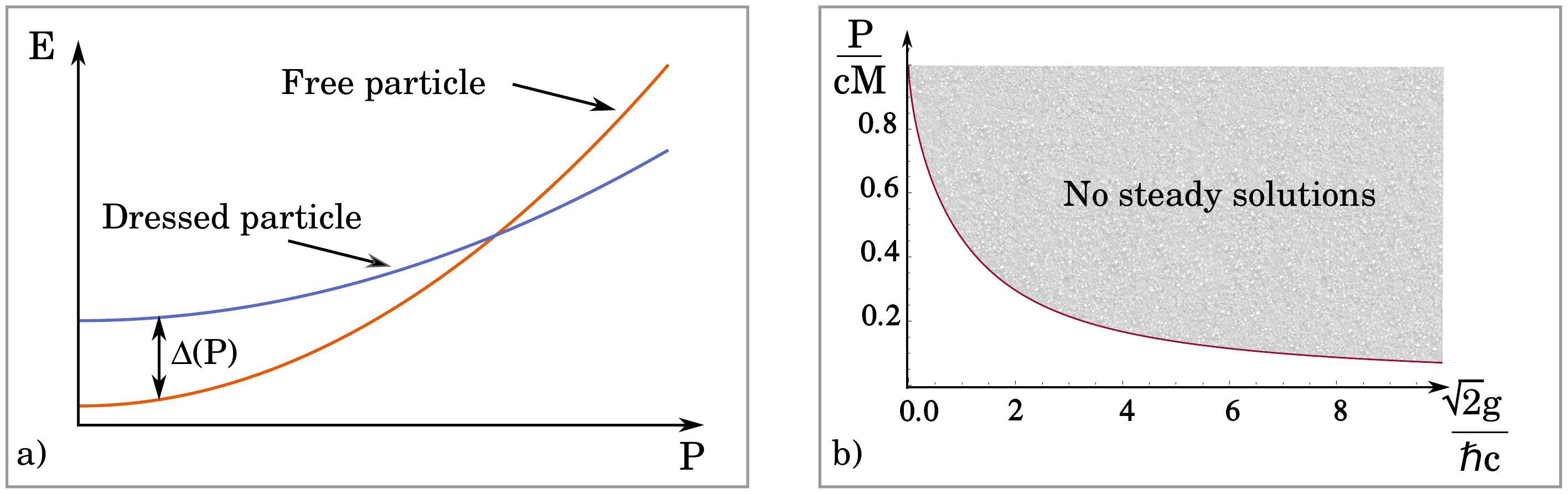}}
\caption{Panel {\bf a)} shows schematically the energies of a free and dressed particles. 
The effective mass and the self-energy 
can be obtained from the energy difference, $\Delta(p)$.
Panel {\bf b)} shows $P_c$ for an impurity of mass $M\gg m$; 
$c$ is the speed of sound in the gas and $g$ is the boson-impurity interaction strength.
  }
\label{fig:Figure3}
\end{figure}
%%%%%%%%%%%%%%%%%%%%%%%%%%%%%%%%%%%%%%%%%%%%%%%%%%%%%%%%%%%%%%%%%%%%%%%%%%%%%%%%%%%%%%%%%%%%%%%%%%%%%%%%%%%%%%%%%%%%%%%

To measure $\epsilon, m_{\mathrm{eff}}$ and $P_c$, one can use a narrow band-pass filter as shown in Fig.~\ref{fig:method_illustration} to create a flux of particles with momenta close to $P$. 
The width (value of $b$) of the target profile must be chosen such that the current is weak, i.e., there is a negligible probability to find two flux particles at distances smaller than the healing 
length of the Bose gas. Then the impurity-in-a-gas picture is applicable by construction. 
For simplicity, we assume that initially the impurity is in a hyperfine state 
that does not interact with the Bose gas. To transfer to a strongly interacting hyperfine state 
one has to deposit enough energy to compensate for the interaction effects; see Fig.~\ref{fig:Figure3}{\bf a)}. 
Therefore, the radio-frequency responce (e.g., the transferred fraction) at different momenta directly measures 
the self-energy and effective mass of the polaron.

A measurement such as this would be similar to the measurement in a recent experiment with a three-dimensional Fermi polaron~\cite{zaccanti2017} but 
with a superior control over the impurity momentum. 
Moreover, the overlap between the non-interacting and interacting states, i.e., the residue, can
be measured, allowing one to test different theoretical methods~\cite{volosniev2017, pastukhov2017,grusdt2017,pastukhov2018}
that, while qualitatively agreeing on $m_{\mathrm{eff}}$ and $\epsilon$, contradict each other on the residue.
Since the momentum of the impurity is known, not only the effective parameters but also the limits of applicability 
of the polaron model will be seen in the radio-frequency responce, in particular, $P_c$.
In Fig.~\ref{fig:Figure3}{\bf b)} we present $P_c$ for impurities whose mass $M$ is much larger than $m$~\cite{hakim1997}.
For weak interactions ($g\to 0$) the critical momentum is determined by the speed of sound, $c$, in accordance with the Landau criterion.
In the opposite limit, $g\to \infty$, the critical momentum goes to zero as $1/g$ (cf.~\cite{kamenev2016}):   
$P_c$ is limited by the timescale for a two-body exchange.

{\it Summary. --}  We outline a procedure for engineering beams of particles with desired momentum profiles using a filter potential connected to a reservoir (see Fig.~\ref{fig:Figure1}). Such a beam can be used to probe cold-atom systems. It can also be used for quantum simulations, as we illustrated with a narrow band-pass filter and a one-dimensional Bose gas in the flux region. Polarons in two-, three- and mixed-dimensional geometries can similarly be created. 

\vspace*{1em}

\begin{acknowledgments}
We thank Peter Schlagheck for referring to~\cite{Schlagheck2005},
and Joachim Brand and Volodymyr Pastukhov for useful discussions.
A.~G.~V. gratefully acknowledges the support of the Humboldt Foundation and the Deutsche Forschungsgemeinschaft
(VO 2437/1-1).

\vspace*{1em}

The authors contributed equally to this work.
\end{acknowledgments}

\bibliographystyle{apsrev4-1}
\bibliography{bib}

\begin{thebibliography}{52}%
\makeatletter
\providecommand \@ifxundefined [1]{%
 \@ifx{#1\undefined}
}%
\providecommand \@ifnum [1]{%
 \ifnum #1\expandafter \@firstoftwo
 \else \expandafter \@secondoftwo
 \fi
}%
\providecommand \@ifx [1]{%
 \ifx #1\expandafter \@firstoftwo
 \else \expandafter \@secondoftwo
 \fi
}%
\providecommand \natexlab [1]{#1}%
\providecommand \enquote  [1]{``#1''}%
\providecommand \bibnamefont  [1]{#1}%
\providecommand \bibfnamefont [1]{#1}%
\providecommand \citenamefont [1]{#1}%
\providecommand \href@noop [0]{\@secondoftwo}%
\providecommand \href [0]{\begingroup \@sanitize@url \@href}%
\providecommand \@href[1]{\@@startlink{#1}\@@href}%
\providecommand \@@href[1]{\endgroup#1\@@endlink}%
\providecommand \@sanitize@url [0]{\catcode `\\12\catcode `\$12\catcode
  `\&12\catcode `\#12\catcode `\^12\catcode `\_12\catcode `\%12\relax}%
\providecommand \@@startlink[1]{}%
\providecommand \@@endlink[0]{}%
\providecommand \url  [0]{\begingroup\@sanitize@url \@url }%
\providecommand \@url [1]{\endgroup\@href {#1}{\urlprefix }}%
\providecommand \urlprefix  [0]{URL }%
\providecommand \Eprint [0]{\href }%
\providecommand \doibase [0]{http://dx.doi.org/}%
\providecommand \selectlanguage [0]{\@gobble}%
\providecommand \bibinfo  [0]{\@secondoftwo}%
\providecommand \bibfield  [0]{\@secondoftwo}%
\providecommand \translation [1]{[#1]}%
\providecommand \BibitemOpen [0]{}%
\providecommand \bibitemStop [0]{}%
\providecommand \bibitemNoStop [0]{.\EOS\space}%
\providecommand \EOS [0]{\spacefactor3000\relax}%
\providecommand \BibitemShut  [1]{\csname bibitem#1\endcsname}%
\let\auto@bib@innerbib\@empty
%</preamble>
\bibitem [{\citenamefont {Braaten}\ and\ \citenamefont
  {Hammer}(2006)}]{braaten2006}%
  \BibitemOpen
  \bibfield  {author} {\bibinfo {author} {\bibfnamefont {E.}~\bibnamefont
  {Braaten}}\ and\ \bibinfo {author} {\bibfnamefont {H.-W.}\ \bibnamefont
  {Hammer}},\ }\href {\doibase 10.1016/j.physrep.2006.03.001} {\bibfield
  {journal} {\bibinfo  {journal} {Phys. Rept.}\ }\textbf {\bibinfo {volume}
  {428}},\ \bibinfo {pages} {259} (\bibinfo {year} {2006})}\BibitemShut
  {NoStop}%
\bibitem [{\citenamefont {Bloch}\ \emph {et~al.}(2008)\citenamefont {Bloch},
  \citenamefont {Dalibard},\ and\ \citenamefont {Zwerger}}]{bloch2008}%
  \BibitemOpen
  \bibfield  {author} {\bibinfo {author} {\bibfnamefont {I.}~\bibnamefont
  {Bloch}}, \bibinfo {author} {\bibfnamefont {J.}~\bibnamefont {Dalibard}}, \
  and\ \bibinfo {author} {\bibfnamefont {W.}~\bibnamefont {Zwerger}},\ }\href
  {\doibase 10.1103/RevModPhys.80.885} {\bibfield  {journal} {\bibinfo
  {journal} {Rev. Mod. Phys.}\ }\textbf {\bibinfo {volume} {80}},\ \bibinfo
  {pages} {885} (\bibinfo {year} {2008})}\BibitemShut {NoStop}%
\bibitem [{\citenamefont {Massignan}\ \emph {et~al.}(2014)\citenamefont
  {Massignan}, \citenamefont {Zaccanti},\ and\ \citenamefont
  {Bruun}}]{massignan2014}%
  \BibitemOpen
  \bibfield  {author} {\bibinfo {author} {\bibfnamefont {P.}~\bibnamefont
  {Massignan}}, \bibinfo {author} {\bibfnamefont {M.}~\bibnamefont {Zaccanti}},
  \ and\ \bibinfo {author} {\bibfnamefont {G.~M.}\ \bibnamefont {Bruun}},\
  }\href {http://stacks.iop.org/0034-4885/77/i=3/a=034401} {\bibfield
  {journal} {\bibinfo  {journal} {Reports on Progress in Physics}\ }\textbf
  {\bibinfo {volume} {77}},\ \bibinfo {pages} {034401} (\bibinfo {year}
  {2014})}\BibitemShut {NoStop}%
\bibitem [{\citenamefont {Schmidt}\ \emph {et~al.}(2018)\citenamefont
  {Schmidt}, \citenamefont {Knap}, \citenamefont {Ivanov}, \citenamefont {You},
  \citenamefont {Cetina},\ and\ \citenamefont {Demler}}]{schmidt2018}%
  \BibitemOpen
  \bibfield  {author} {\bibinfo {author} {\bibfnamefont {R.}~\bibnamefont
  {Schmidt}}, \bibinfo {author} {\bibfnamefont {M.}~\bibnamefont {Knap}},
  \bibinfo {author} {\bibfnamefont {D.~A.}\ \bibnamefont {Ivanov}}, \bibinfo
  {author} {\bibfnamefont {J.-S.}\ \bibnamefont {You}}, \bibinfo {author}
  {\bibfnamefont {M.}~\bibnamefont {Cetina}}, \ and\ \bibinfo {author}
  {\bibfnamefont {E.}~\bibnamefont {Demler}},\ }\href
  {http://iopscience.iop.org/article/10.1088/1361-6633/aa9593/meta} {\bibfield
  {journal} {\bibinfo  {journal} {Rep. Prog. Phys.}\ }\textbf {\bibinfo
  {volume} {81}},\ \bibinfo {pages} {024401} (\bibinfo {year}
  {2018})}\BibitemShut {NoStop}%
\bibitem [{\citenamefont {Impens}\ \emph {et~al.}(2006)\citenamefont {Impens},
  \citenamefont {Bouyer},\ and\ \citenamefont {Bord{\'e}}}]{Impens2006}%
  \BibitemOpen
  \bibfield  {author} {\bibinfo {author} {\bibfnamefont {F.}~\bibnamefont
  {Impens}}, \bibinfo {author} {\bibfnamefont {P.}~\bibnamefont {Bouyer}}, \
  and\ \bibinfo {author} {\bibfnamefont {C.}~\bibnamefont {Bord{\'e}}},\ }\href
  {\doibase 10.1007/s00340-006-2399-3} {\bibfield  {journal} {\bibinfo
  {journal} {Applied Physics B}\ }\textbf {\bibinfo {volume} {84}},\ \bibinfo
  {pages} {603} (\bibinfo {year} {2006})}\BibitemShut {NoStop}%
\bibitem [{\citenamefont {Cronin}\ \emph {et~al.}(2009)\citenamefont {Cronin},
  \citenamefont {Schmiedmayer},\ and\ \citenamefont {Pritchard}}]{Cronin2009}%
  \BibitemOpen
  \bibfield  {author} {\bibinfo {author} {\bibfnamefont {A.~D.}\ \bibnamefont
  {Cronin}}, \bibinfo {author} {\bibfnamefont {J.}~\bibnamefont
  {Schmiedmayer}}, \ and\ \bibinfo {author} {\bibfnamefont {D.~E.}\
  \bibnamefont {Pritchard}},\ }\href {\doibase 10.1103/RevModPhys.81.1051}
  {\bibfield  {journal} {\bibinfo  {journal} {Rev. Mod. Phys.}\ }\textbf
  {\bibinfo {volume} {81}},\ \bibinfo {pages} {1051} (\bibinfo {year}
  {2009})}\BibitemShut {NoStop}%
\bibitem [{\citenamefont {Micheli}\ \emph {et~al.}(2004)\citenamefont
  {Micheli}, \citenamefont {Daley}, \citenamefont {Jaksch},\ and\ \citenamefont
  {Zoller}}]{zoller2004}%
  \BibitemOpen
  \bibfield  {author} {\bibinfo {author} {\bibfnamefont {A.}~\bibnamefont
  {Micheli}}, \bibinfo {author} {\bibfnamefont {A.~J.}\ \bibnamefont {Daley}},
  \bibinfo {author} {\bibfnamefont {D.}~\bibnamefont {Jaksch}}, \ and\ \bibinfo
  {author} {\bibfnamefont {P.}~\bibnamefont {Zoller}},\ }\href {\doibase
  10.1103/PhysRevLett.93.140408} {\bibfield  {journal} {\bibinfo  {journal}
  {Phys. Rev. Lett.}\ }\textbf {\bibinfo {volume} {93}},\ \bibinfo {pages}
  {140408} (\bibinfo {year} {2004})}\BibitemShut {NoStop}%
\bibitem [{\citenamefont {Marchukov}\ \emph {et~al.}(2016)\citenamefont
  {Marchukov}, \citenamefont {Volosniev}, \citenamefont {Valiente},
  \citenamefont {Petrosyan},\ and\ \citenamefont {Zinner}}]{marchukov2016}%
  \BibitemOpen
  \bibfield  {author} {\bibinfo {author} {\bibfnamefont {O.~V.}\ \bibnamefont
  {Marchukov}}, \bibinfo {author} {\bibfnamefont {A.~G.}\ \bibnamefont
  {Volosniev}}, \bibinfo {author} {\bibfnamefont {M.}~\bibnamefont {Valiente}},
  \bibinfo {author} {\bibfnamefont {D.}~\bibnamefont {Petrosyan}}, \ and\
  \bibinfo {author} {\bibfnamefont {N.~T.}\ \bibnamefont {Zinner}},\ }\href
  {\doibase 10.1038/ncomms13070} {\bibfield  {journal} {\bibinfo  {journal}
  {Nature Communications}\ }\textbf {\bibinfo {volume} {7}},\ \bibinfo {pages}
  {13070} (\bibinfo {year} {2016})}\BibitemShut {NoStop}%
\bibitem [{\citenamefont {Thuberg}\ \emph {et~al.}(2017)\citenamefont
  {Thuberg}, \citenamefont {Mu\~noz}, \citenamefont {Eggert},\ and\
  \citenamefont {Reyes}}]{thuberg2017}%
  \BibitemOpen
  \bibfield  {author} {\bibinfo {author} {\bibfnamefont {D.}~\bibnamefont
  {Thuberg}}, \bibinfo {author} {\bibfnamefont {E.}~\bibnamefont {Mu\~noz}},
  \bibinfo {author} {\bibfnamefont {S.}~\bibnamefont {Eggert}}, \ and\ \bibinfo
  {author} {\bibfnamefont {S.~A.}\ \bibnamefont {Reyes}},\ }\href {\doibase
  10.1103/PhysRevLett.119.267701} {\bibfield  {journal} {\bibinfo  {journal}
  {Phys. Rev. Lett.}\ }\textbf {\bibinfo {volume} {119}},\ \bibinfo {pages}
  {267701} (\bibinfo {year} {2017})}\BibitemShut {NoStop}%
\bibitem [{Note1()}]{Note1}%
  \BibitemOpen
  \bibinfo {note} {Strong interactions can lead to the transmission behavior
  that is not captured by the one-body Schr{\"o}dinger equation, e.g.,to a
  collective resonant transport~\cite {Schlagheck2005}.}\BibitemShut {Stop}%
\bibitem [{\citenamefont {Krinner}\ \emph {et~al.}(2015)\citenamefont
  {Krinner}, \citenamefont {Husmann}, \citenamefont {Brantut},\ and\
  \citenamefont {Esslinger}}]{esslinger2015}%
  \BibitemOpen
  \bibfield  {author} {\bibinfo {author} {\bibfnamefont {S.}~\bibnamefont
  {Krinner}}, \bibinfo {author} {\bibfnamefont {D.}~\bibnamefont {Husmann}},
  \bibinfo {author} {\bibfnamefont {J.-P.}\ \bibnamefont {Brantut}}, \ and\
  \bibinfo {author} {\bibfnamefont {T.}~\bibnamefont {Esslinger}},\ }\href
  {\doibase 10.1038/nature14049} {\bibfield  {journal} {\bibinfo  {journal}
  {Nature}\ }\textbf {\bibinfo {volume} {517}},\ \bibinfo {pages} {64}
  (\bibinfo {year} {2015})}\BibitemShut {NoStop}%
\bibitem [{\citenamefont {Lebrat}\ \emph {et~al.}(2018)\citenamefont {Lebrat},
  \citenamefont {Gri\ifmmode~\check{s}\else \v{s}\fi{}ins}, \citenamefont
  {Husmann}, \citenamefont {H\"ausler}, \citenamefont {Corman}, \citenamefont
  {Giamarchi}, \citenamefont {Brantut},\ and\ \citenamefont
  {Esslinger}}]{esslinger2018}%
  \BibitemOpen
  \bibfield  {author} {\bibinfo {author} {\bibfnamefont {M.}~\bibnamefont
  {Lebrat}}, \bibinfo {author} {\bibfnamefont {P.}~\bibnamefont
  {Gri\ifmmode~\check{s}\else \v{s}\fi{}ins}}, \bibinfo {author} {\bibfnamefont
  {D.}~\bibnamefont {Husmann}}, \bibinfo {author} {\bibfnamefont
  {S.}~\bibnamefont {H\"ausler}}, \bibinfo {author} {\bibfnamefont
  {L.}~\bibnamefont {Corman}}, \bibinfo {author} {\bibfnamefont
  {T.}~\bibnamefont {Giamarchi}}, \bibinfo {author} {\bibfnamefont {J.-P.}\
  \bibnamefont {Brantut}}, \ and\ \bibinfo {author} {\bibfnamefont
  {T.}~\bibnamefont {Esslinger}},\ }\href {\doibase 10.1103/PhysRevX.8.011053}
  {\bibfield  {journal} {\bibinfo  {journal} {Phys. Rev. X}\ }\textbf {\bibinfo
  {volume} {8}},\ \bibinfo {pages} {011053} (\bibinfo {year}
  {2018})}\BibitemShut {NoStop}%
\bibitem [{\citenamefont {Lebrat}\ \emph {et~al.}(2019)\citenamefont {Lebrat},
  \citenamefont {H\"ausler}, \citenamefont {Fabritius}, \citenamefont
  {Husmann}, \citenamefont {Corman},\ and\ \citenamefont
  {Esslinger}}]{esslinger2019}%
  \BibitemOpen
  \bibfield  {author} {\bibinfo {author} {\bibfnamefont {M.}~\bibnamefont
  {Lebrat}}, \bibinfo {author} {\bibfnamefont {S.}~\bibnamefont {H\"ausler}},
  \bibinfo {author} {\bibfnamefont {P.}~\bibnamefont {Fabritius}}, \bibinfo
  {author} {\bibfnamefont {D.}~\bibnamefont {Husmann}}, \bibinfo {author}
  {\bibfnamefont {L.}~\bibnamefont {Corman}}, \ and\ \bibinfo {author}
  {\bibfnamefont {T.}~\bibnamefont {Esslinger}},\ }\href@noop {} {\bibfield
  {journal} {\bibinfo  {journal} {arXiv:{\bf 1902.05516}}\ } (\bibinfo {year}
  {2019})}\BibitemShut {NoStop}%
\bibitem [{\citenamefont {Bergschneider}\ \emph {et~al.}(2018)\citenamefont
  {Bergschneider}, \citenamefont {Klinkhamer}, \citenamefont {Becher},
  \citenamefont {Klemt}, \citenamefont {Z\"urn}, \citenamefont {Preiss},\ and\
  \citenamefont {Jochim}}]{jochim2018}%
  \BibitemOpen
  \bibfield  {author} {\bibinfo {author} {\bibfnamefont {A.}~\bibnamefont
  {Bergschneider}}, \bibinfo {author} {\bibfnamefont {V.~M.}\ \bibnamefont
  {Klinkhamer}}, \bibinfo {author} {\bibfnamefont {J.~H.}\ \bibnamefont
  {Becher}}, \bibinfo {author} {\bibfnamefont {R.}~\bibnamefont {Klemt}},
  \bibinfo {author} {\bibfnamefont {G.}~\bibnamefont {Z\"urn}}, \bibinfo
  {author} {\bibfnamefont {P.~M.}\ \bibnamefont {Preiss}}, \ and\ \bibinfo
  {author} {\bibfnamefont {S.}~\bibnamefont {Jochim}},\ }\href {\doibase
  10.1103/PhysRevA.97.063613} {\bibfield  {journal} {\bibinfo  {journal} {Phys.
  Rev. A}\ }\textbf {\bibinfo {volume} {97}},\ \bibinfo {pages} {063613}
  (\bibinfo {year} {2018})}\BibitemShut {NoStop}%
\bibitem [{\citenamefont {Guerin}\ \emph {et~al.}(2006)\citenamefont {Guerin},
  \citenamefont {Riou}, \citenamefont {Gaebler}, \citenamefont {Josse},
  \citenamefont {Bouyer},\ and\ \citenamefont
  {Aspect}}]{PhysRevLett.97.200402}%
  \BibitemOpen
  \bibfield  {author} {\bibinfo {author} {\bibfnamefont {W.}~\bibnamefont
  {Guerin}}, \bibinfo {author} {\bibfnamefont {J.-F.}\ \bibnamefont {Riou}},
  \bibinfo {author} {\bibfnamefont {J.~P.}\ \bibnamefont {Gaebler}}, \bibinfo
  {author} {\bibfnamefont {V.}~\bibnamefont {Josse}}, \bibinfo {author}
  {\bibfnamefont {P.}~\bibnamefont {Bouyer}}, \ and\ \bibinfo {author}
  {\bibfnamefont {A.}~\bibnamefont {Aspect}},\ }\href {\doibase
  10.1103/PhysRevLett.97.200402} {\bibfield  {journal} {\bibinfo  {journal}
  {Phys. Rev. Lett.}\ }\textbf {\bibinfo {volume} {97}},\ \bibinfo {pages}
  {200402} (\bibinfo {year} {2006})}\BibitemShut {NoStop}%
\bibitem [{\citenamefont {Gattobigio}\ \emph {et~al.}(2009)\citenamefont
  {Gattobigio}, \citenamefont {Couvert}, \citenamefont {Jeppesen},
  \citenamefont {Mathevet},\ and\ \citenamefont
  {Gu\'ery-Odelin}}]{PhysRevA.80.041605}%
  \BibitemOpen
  \bibfield  {author} {\bibinfo {author} {\bibfnamefont {G.~L.}\ \bibnamefont
  {Gattobigio}}, \bibinfo {author} {\bibfnamefont {A.}~\bibnamefont {Couvert}},
  \bibinfo {author} {\bibfnamefont {M.}~\bibnamefont {Jeppesen}}, \bibinfo
  {author} {\bibfnamefont {R.}~\bibnamefont {Mathevet}}, \ and\ \bibinfo
  {author} {\bibfnamefont {D.}~\bibnamefont {Gu\'ery-Odelin}},\ }\href
  {\doibase 10.1103/PhysRevA.80.041605} {\bibfield  {journal} {\bibinfo
  {journal} {Phys. Rev. A}\ }\textbf {\bibinfo {volume} {80}},\ \bibinfo
  {pages} {041605} (\bibinfo {year} {2009})}\BibitemShut {NoStop}%
\bibitem [{\citenamefont {B{\"u}ning}\ \emph {et~al.}(2010)\citenamefont
  {B{\"u}ning}, \citenamefont {Will}, \citenamefont {Ertmer}, \citenamefont
  {Klempt},\ and\ \citenamefont {Arlt}}]{buning2010slow}%
  \BibitemOpen
  \bibfield  {author} {\bibinfo {author} {\bibfnamefont {G.~K.}\ \bibnamefont
  {B{\"u}ning}}, \bibinfo {author} {\bibfnamefont {J.}~\bibnamefont {Will}},
  \bibinfo {author} {\bibfnamefont {W.}~\bibnamefont {Ertmer}}, \bibinfo
  {author} {\bibfnamefont {C.}~\bibnamefont {Klempt}}, \ and\ \bibinfo {author}
  {\bibfnamefont {J.}~\bibnamefont {Arlt}},\ }\href@noop {} {\bibfield
  {journal} {\bibinfo  {journal} {Applied Physics B}\ }\textbf {\bibinfo
  {volume} {100}},\ \bibinfo {pages} {117} (\bibinfo {year}
  {2010})}\BibitemShut {NoStop}%
\bibitem [{\citenamefont {Carusotto}\ and\ \citenamefont
  {La~Rocca}(2000)}]{PhysRevLett.84.399}%
  \BibitemOpen
  \bibfield  {author} {\bibinfo {author} {\bibfnamefont {I.}~\bibnamefont
  {Carusotto}}\ and\ \bibinfo {author} {\bibfnamefont {G.~C.}\ \bibnamefont
  {La~Rocca}},\ }\href {\doibase 10.1103/PhysRevLett.84.399} {\bibfield
  {journal} {\bibinfo  {journal} {Phys. Rev. Lett.}\ }\textbf {\bibinfo
  {volume} {84}},\ \bibinfo {pages} {399} (\bibinfo {year} {2000})}\BibitemShut
  {NoStop}%
\bibitem [{\citenamefont {Fabre}\ \emph {et~al.}(2011)\citenamefont {Fabre},
  \citenamefont {Cheiney}, \citenamefont {Gattobigio}, \citenamefont
  {Vermersch}, \citenamefont {Faure}, \citenamefont {Mathevet}, \citenamefont
  {Lahaye},\ and\ \citenamefont {Gu\'ery-Odelin}}]{PhysRevLett.107.230401}%
  \BibitemOpen
  \bibfield  {author} {\bibinfo {author} {\bibfnamefont {C.~M.}\ \bibnamefont
  {Fabre}}, \bibinfo {author} {\bibfnamefont {P.}~\bibnamefont {Cheiney}},
  \bibinfo {author} {\bibfnamefont {G.~L.}\ \bibnamefont {Gattobigio}},
  \bibinfo {author} {\bibfnamefont {F.}~\bibnamefont {Vermersch}}, \bibinfo
  {author} {\bibfnamefont {S.}~\bibnamefont {Faure}}, \bibinfo {author}
  {\bibfnamefont {R.}~\bibnamefont {Mathevet}}, \bibinfo {author}
  {\bibfnamefont {T.}~\bibnamefont {Lahaye}}, \ and\ \bibinfo {author}
  {\bibfnamefont {D.}~\bibnamefont {Gu\'ery-Odelin}},\ }\href {\doibase
  10.1103/PhysRevLett.107.230401} {\bibfield  {journal} {\bibinfo  {journal}
  {Phys. Rev. Lett.}\ }\textbf {\bibinfo {volume} {107}},\ \bibinfo {pages}
  {230401} (\bibinfo {year} {2011})}\BibitemShut {NoStop}%
\bibitem [{\citenamefont {Damon}\ \emph {et~al.}(2015)\citenamefont {Damon},
  \citenamefont {Condon}, \citenamefont {Cheiney}, \citenamefont {Fortun},
  \citenamefont {Georgeot}, \citenamefont {Billy},\ and\ \citenamefont
  {Gu\'ery-Odelin}}]{PhysRevA.92.033614}%
  \BibitemOpen
  \bibfield  {author} {\bibinfo {author} {\bibfnamefont {F.}~\bibnamefont
  {Damon}}, \bibinfo {author} {\bibfnamefont {G.}~\bibnamefont {Condon}},
  \bibinfo {author} {\bibfnamefont {P.}~\bibnamefont {Cheiney}}, \bibinfo
  {author} {\bibfnamefont {A.}~\bibnamefont {Fortun}}, \bibinfo {author}
  {\bibfnamefont {B.}~\bibnamefont {Georgeot}}, \bibinfo {author}
  {\bibfnamefont {J.}~\bibnamefont {Billy}}, \ and\ \bibinfo {author}
  {\bibfnamefont {D.}~\bibnamefont {Gu\'ery-Odelin}},\ }\href {\doibase
  10.1103/PhysRevA.92.033614} {\bibfield  {journal} {\bibinfo  {journal} {Phys.
  Rev. A}\ }\textbf {\bibinfo {volume} {92}},\ \bibinfo {pages} {033614}
  (\bibinfo {year} {2015})}\BibitemShut {NoStop}%
\bibitem [{\citenamefont {Landau}\ and\ \citenamefont
  {Lifshitz}(1977)}]{landau1977}%
  \BibitemOpen
  \bibfield  {author} {\bibinfo {author} {\bibfnamefont {L.~D.}\ \bibnamefont
  {Landau}}\ and\ \bibinfo {author} {\bibfnamefont {E.~M.}\ \bibnamefont
  {Lifshitz}},\ }\href@noop {} {\emph {\bibinfo {title} {Quantum Mechanics}}}\
  (\bibinfo  {publisher} {Butterworth-Heinemann, Oxford},\ \bibinfo {year}
  {1977})\BibitemShut {NoStop}%
\bibitem [{\citenamefont {Storn}\ and\ \citenamefont
  {Price}(1997)}]{storn1997differential}%
  \BibitemOpen
  \bibfield  {author} {\bibinfo {author} {\bibfnamefont {R.}~\bibnamefont
  {Storn}}\ and\ \bibinfo {author} {\bibfnamefont {K.}~\bibnamefont {Price}},\
  }\href@noop {} {\bibfield  {journal} {\bibinfo  {journal} {Journal of global
  optimization}\ }\textbf {\bibinfo {volume} {11}},\ \bibinfo {pages} {341}
  (\bibinfo {year} {1997})}\BibitemShut {NoStop}%
\bibitem [{Note2()}]{Note2}%
  \BibitemOpen
  \bibinfo {note} {For a reservoir at a finite temperature it is beneficial to
  include the energy distribution, $n(k)$, in the cost function. To this end,
  one should simply convolute $T_{\protect \bm {\theta }}(k)$ in Eq.~(\ref
  {eq:cost1}) with $n(k)$. The function $n(k)$ is determined by the
  Bose-Einstein or Fermi-Dirac distributions, temperature and the shape of the
  trap. For simplicity, we assume in current Eq.~(\ref {eq:cost1}) a
  one-dimensional Fermi gas at zero temperature in a harmonic trap, i.e.,
  $n(k<k_F)=1$ and zero otherwise.}\BibitemShut {Stop}%
\bibitem [{\citenamefont {Schirotzek}\ \emph {et~al.}(2009)\citenamefont
  {Schirotzek}, \citenamefont {Wu}, \citenamefont {Sommer},\ and\ \citenamefont
  {Zwierlein}}]{zwierlein2009}%
  \BibitemOpen
  \bibfield  {author} {\bibinfo {author} {\bibfnamefont {A.}~\bibnamefont
  {Schirotzek}}, \bibinfo {author} {\bibfnamefont {C.-H.}\ \bibnamefont {Wu}},
  \bibinfo {author} {\bibfnamefont {A.}~\bibnamefont {Sommer}}, \ and\ \bibinfo
  {author} {\bibfnamefont {M.~W.}\ \bibnamefont {Zwierlein}},\ }\href {\doibase
  10.1103/PhysRevLett.102.230402} {\bibfield  {journal} {\bibinfo  {journal}
  {Phys. Rev. Lett.}\ }\textbf {\bibinfo {volume} {102}},\ \bibinfo {pages}
  {230402} (\bibinfo {year} {2009})}\BibitemShut {NoStop}%
\bibitem [{\citenamefont {Nascimb\`ene}\ \emph {et~al.}(2009)\citenamefont
  {Nascimb\`ene}, \citenamefont {Navon}, \citenamefont {Jiang}, \citenamefont
  {Tarruell}, \citenamefont {Teichmann}, \citenamefont {McKeever},
  \citenamefont {Chevy},\ and\ \citenamefont {Salomon}}]{salomon2009}%
  \BibitemOpen
  \bibfield  {author} {\bibinfo {author} {\bibfnamefont {S.}~\bibnamefont
  {Nascimb\`ene}}, \bibinfo {author} {\bibfnamefont {N.}~\bibnamefont {Navon}},
  \bibinfo {author} {\bibfnamefont {K.~J.}\ \bibnamefont {Jiang}}, \bibinfo
  {author} {\bibfnamefont {L.}~\bibnamefont {Tarruell}}, \bibinfo {author}
  {\bibfnamefont {M.}~\bibnamefont {Teichmann}}, \bibinfo {author}
  {\bibfnamefont {J.}~\bibnamefont {McKeever}}, \bibinfo {author}
  {\bibfnamefont {F.}~\bibnamefont {Chevy}}, \ and\ \bibinfo {author}
  {\bibfnamefont {C.}~\bibnamefont {Salomon}},\ }\href {\doibase
  10.1103/PhysRevLett.103.170402} {\bibfield  {journal} {\bibinfo  {journal}
  {Phys. Rev. Lett.}\ }\textbf {\bibinfo {volume} {103}},\ \bibinfo {pages}
  {170402} (\bibinfo {year} {2009})}\BibitemShut {NoStop}%
\bibitem [{\citenamefont {Kohstall}\ \emph {et~al.}(2012)\citenamefont
  {Kohstall}, \citenamefont {Zaccanti}, \citenamefont {Jag}, \citenamefont
  {Trenkwalder}, \citenamefont {P.Massignan}, \citenamefont {Bruun},
  \citenamefont {Schreck},\ and\ \citenamefont {Grimm}}]{grimm2012}%
  \BibitemOpen
  \bibfield  {author} {\bibinfo {author} {\bibfnamefont {C.}~\bibnamefont
  {Kohstall}}, \bibinfo {author} {\bibfnamefont {M.}~\bibnamefont {Zaccanti}},
  \bibinfo {author} {\bibfnamefont {M.}~\bibnamefont {Jag}}, \bibinfo {author}
  {\bibfnamefont {A.}~\bibnamefont {Trenkwalder}}, \bibinfo {author}
  {\bibnamefont {P.Massignan}}, \bibinfo {author} {\bibfnamefont {G.~M.}\
  \bibnamefont {Bruun}}, \bibinfo {author} {\bibfnamefont {F.}~\bibnamefont
  {Schreck}}, \ and\ \bibinfo {author} {\bibfnamefont {R.}~\bibnamefont
  {Grimm}},\ }\href@noop {} {\bibfield  {journal} {\bibinfo  {journal}
  {Nature}\ }\textbf {\bibinfo {volume} {485}},\ \bibinfo {pages} {615}
  (\bibinfo {year} {2012})}\BibitemShut {NoStop}%
\bibitem [{\citenamefont {Spethmann}\ \emph {et~al.}(2012)\citenamefont
  {Spethmann}, \citenamefont {Kindermann}, \citenamefont {John}, \citenamefont
  {Weber}, \citenamefont {Meschede},\ and\ \citenamefont
  {Widera}}]{widera2012}%
  \BibitemOpen
  \bibfield  {author} {\bibinfo {author} {\bibfnamefont {N.}~\bibnamefont
  {Spethmann}}, \bibinfo {author} {\bibfnamefont {F.}~\bibnamefont
  {Kindermann}}, \bibinfo {author} {\bibfnamefont {S.}~\bibnamefont {John}},
  \bibinfo {author} {\bibfnamefont {C.}~\bibnamefont {Weber}}, \bibinfo
  {author} {\bibfnamefont {D.}~\bibnamefont {Meschede}}, \ and\ \bibinfo
  {author} {\bibfnamefont {A.}~\bibnamefont {Widera}},\ }\href {\doibase
  10.1103/PhysRevLett.109.235301} {\bibfield  {journal} {\bibinfo  {journal}
  {Phys. Rev. Lett.}\ }\textbf {\bibinfo {volume} {109}},\ \bibinfo {pages}
  {235301} (\bibinfo {year} {2012})}\BibitemShut {NoStop}%
\bibitem [{\citenamefont {Catani}\ \emph {et~al.}(2012)\citenamefont {Catani},
  \citenamefont {Lamporesi}, \citenamefont {Naik}, \citenamefont {Gring},
  \citenamefont {Inguscio}, \citenamefont {Minardi}, \citenamefont {Kantian},\
  and\ \citenamefont {Giamarchi}}]{catani2012}%
  \BibitemOpen
  \bibfield  {author} {\bibinfo {author} {\bibfnamefont {J.}~\bibnamefont
  {Catani}}, \bibinfo {author} {\bibfnamefont {G.}~\bibnamefont {Lamporesi}},
  \bibinfo {author} {\bibfnamefont {D.}~\bibnamefont {Naik}}, \bibinfo {author}
  {\bibfnamefont {M.}~\bibnamefont {Gring}}, \bibinfo {author} {\bibfnamefont
  {M.}~\bibnamefont {Inguscio}}, \bibinfo {author} {\bibfnamefont
  {F.}~\bibnamefont {Minardi}}, \bibinfo {author} {\bibfnamefont
  {A.}~\bibnamefont {Kantian}}, \ and\ \bibinfo {author} {\bibfnamefont
  {T.}~\bibnamefont {Giamarchi}},\ }\href {\doibase 10.1103/PhysRevA.85.023623}
  {\bibfield  {journal} {\bibinfo  {journal} {Phys. Rev. A}\ }\textbf {\bibinfo
  {volume} {85}},\ \bibinfo {pages} {023623} (\bibinfo {year}
  {2012})}\BibitemShut {NoStop}%
\bibitem [{\citenamefont {Fukuhara}\ \emph {et~al.}(2013)\citenamefont
  {Fukuhara}, \citenamefont {Kantian}, \citenamefont {Endres}, \citenamefont
  {Cheneau}, \citenamefont {Schaulss}, \citenamefont {Hild}, \citenamefont
  {Bellem}, \citenamefont {Schollw{\"o}ck}, \citenamefont {Giamarchi},
  \citenamefont {Gross}, \citenamefont {Bloch},\ and\ \citenamefont
  {Kuhr}}]{fukuhara2013}%
  \BibitemOpen
  \bibfield  {author} {\bibinfo {author} {\bibfnamefont {T.}~\bibnamefont
  {Fukuhara}}, \bibinfo {author} {\bibfnamefont {A.}~\bibnamefont {Kantian}},
  \bibinfo {author} {\bibfnamefont {M.}~\bibnamefont {Endres}}, \bibinfo
  {author} {\bibfnamefont {M.}~\bibnamefont {Cheneau}}, \bibinfo {author}
  {\bibfnamefont {P.}~\bibnamefont {Schaulss}}, \bibinfo {author}
  {\bibfnamefont {S.}~\bibnamefont {Hild}}, \bibinfo {author} {\bibfnamefont
  {D.}~\bibnamefont {Bellem}}, \bibinfo {author} {\bibfnamefont
  {U.}~\bibnamefont {Schollw{\"o}ck}}, \bibinfo {author} {\bibfnamefont
  {T.}~\bibnamefont {Giamarchi}}, \bibinfo {author} {\bibfnamefont
  {C.}~\bibnamefont {Gross}}, \bibinfo {author} {\bibfnamefont
  {I.}~\bibnamefont {Bloch}}, \ and\ \bibinfo {author} {\bibfnamefont
  {S.}~\bibnamefont {Kuhr}},\ }\href@noop {} {\bibfield  {journal} {\bibinfo
  {journal} {Nature Physics}\ }\textbf {\bibinfo {volume} {9}},\ \bibinfo
  {pages} {235} (\bibinfo {year} {2013})}\BibitemShut {NoStop}%
\bibitem [{\citenamefont {Hu}\ \emph {et~al.}(2016)\citenamefont {Hu},
  \citenamefont {Van~de Graaff}, \citenamefont {Kedar}, \citenamefont {Corson},
  \citenamefont {Cornell},\ and\ \citenamefont {Jin}}]{hu2016}%
  \BibitemOpen
  \bibfield  {author} {\bibinfo {author} {\bibfnamefont {M.-G.}\ \bibnamefont
  {Hu}}, \bibinfo {author} {\bibfnamefont {M.~J.}\ \bibnamefont {Van~de
  Graaff}}, \bibinfo {author} {\bibfnamefont {D.}~\bibnamefont {Kedar}},
  \bibinfo {author} {\bibfnamefont {J.~P.}\ \bibnamefont {Corson}}, \bibinfo
  {author} {\bibfnamefont {E.~A.}\ \bibnamefont {Cornell}}, \ and\ \bibinfo
  {author} {\bibfnamefont {D.~S.}\ \bibnamefont {Jin}},\ }\href {\doibase
  10.1103/PhysRevLett.117.055301} {\bibfield  {journal} {\bibinfo  {journal}
  {Phys. Rev. Lett.}\ }\textbf {\bibinfo {volume} {117}},\ \bibinfo {pages}
  {055301} (\bibinfo {year} {2016})}\BibitemShut {NoStop}%
\bibitem [{\citenamefont {J\o{}rgensen}\ \emph {et~al.}(2016)\citenamefont
  {J\o{}rgensen}, \citenamefont {Wacker}, \citenamefont {Skalmstang},
  \citenamefont {Parish}, \citenamefont {Levinsen}, \citenamefont
  {Christensen}, \citenamefont {Bruun},\ and\ \citenamefont {Arlt}}]{arlt2016}%
  \BibitemOpen
  \bibfield  {author} {\bibinfo {author} {\bibfnamefont {N.~B.}\ \bibnamefont
  {J\o{}rgensen}}, \bibinfo {author} {\bibfnamefont {L.}~\bibnamefont
  {Wacker}}, \bibinfo {author} {\bibfnamefont {K.~T.}\ \bibnamefont
  {Skalmstang}}, \bibinfo {author} {\bibfnamefont {M.~M.}\ \bibnamefont
  {Parish}}, \bibinfo {author} {\bibfnamefont {J.}~\bibnamefont {Levinsen}},
  \bibinfo {author} {\bibfnamefont {R.~S.}\ \bibnamefont {Christensen}},
  \bibinfo {author} {\bibfnamefont {G.~M.}\ \bibnamefont {Bruun}}, \ and\
  \bibinfo {author} {\bibfnamefont {J.~J.}\ \bibnamefont {Arlt}},\ }\href
  {\doibase 10.1103/PhysRevLett.117.055302} {\bibfield  {journal} {\bibinfo
  {journal} {Phys. Rev. Lett.}\ }\textbf {\bibinfo {volume} {117}},\ \bibinfo
  {pages} {055302} (\bibinfo {year} {2016})}\BibitemShut {NoStop}%
\bibitem [{\citenamefont {Scazza}\ \emph {et~al.}(2017)\citenamefont {Scazza},
  \citenamefont {Valtolina}, \citenamefont {Massignan}, \citenamefont {Recati},
  \citenamefont {Amico}, \citenamefont {Burchianti}, \citenamefont {Fort},
  \citenamefont {Inguscio}, \citenamefont {Zaccanti},\ and\ \citenamefont
  {Roati}}]{zaccanti2017}%
  \BibitemOpen
  \bibfield  {author} {\bibinfo {author} {\bibfnamefont {F.}~\bibnamefont
  {Scazza}}, \bibinfo {author} {\bibfnamefont {G.}~\bibnamefont {Valtolina}},
  \bibinfo {author} {\bibfnamefont {P.}~\bibnamefont {Massignan}}, \bibinfo
  {author} {\bibfnamefont {A.}~\bibnamefont {Recati}}, \bibinfo {author}
  {\bibfnamefont {A.}~\bibnamefont {Amico}}, \bibinfo {author} {\bibfnamefont
  {A.}~\bibnamefont {Burchianti}}, \bibinfo {author} {\bibfnamefont
  {C.}~\bibnamefont {Fort}}, \bibinfo {author} {\bibfnamefont {M.}~\bibnamefont
  {Inguscio}}, \bibinfo {author} {\bibfnamefont {M.}~\bibnamefont {Zaccanti}},
  \ and\ \bibinfo {author} {\bibfnamefont {G.}~\bibnamefont {Roati}},\ }\href
  {\doibase 10.1103/PhysRevLett.118.083602} {\bibfield  {journal} {\bibinfo
  {journal} {Phys. Rev. Lett.}\ }\textbf {\bibinfo {volume} {118}},\ \bibinfo
  {pages} {083602} (\bibinfo {year} {2017})}\BibitemShut {NoStop}%
\bibitem [{\citenamefont {Pe{\~n}a~Ardila}\ \emph {et~al.}(2019)\citenamefont
  {Pe{\~n}a~Ardila}, \citenamefont {J{\"o}rgensen}, \citenamefont {Pohl},
  \citenamefont {Giorgini}, \citenamefont {Bruun},\ and\ \citenamefont
  {Arlt}}]{ardila2018}%
  \BibitemOpen
  \bibfield  {author} {\bibinfo {author} {\bibfnamefont {L.~A.}\ \bibnamefont
  {Pe{\~n}a~Ardila}}, \bibinfo {author} {\bibfnamefont {N.~B.}\ \bibnamefont
  {J{\"o}rgensen}}, \bibinfo {author} {\bibfnamefont {T.}~\bibnamefont {Pohl}},
  \bibinfo {author} {\bibfnamefont {S.}~\bibnamefont {Giorgini}}, \bibinfo
  {author} {\bibfnamefont {G.~M.}\ \bibnamefont {Bruun}}, \ and\ \bibinfo
  {author} {\bibfnamefont {J.~J.}\ \bibnamefont {Arlt}},\ }\href@noop {}
  {\bibfield  {journal} {\bibinfo  {journal} {Phys. Rev. A}\ }\textbf {\bibinfo
  {volume} {99}},\ \bibinfo {pages} {063607} (\bibinfo {year}
  {2019})}\BibitemShut {NoStop}%
\bibitem [{\citenamefont {Hakim}(1997)}]{hakim1997}%
  \BibitemOpen
  \bibfield  {author} {\bibinfo {author} {\bibfnamefont {V.}~\bibnamefont
  {Hakim}},\ }\href {\doibase 10.1103/PhysRevE.55.2835} {\bibfield  {journal}
  {\bibinfo  {journal} {Phys. Rev. E}\ }\textbf {\bibinfo {volume} {55}},\
  \bibinfo {pages} {2835} (\bibinfo {year} {1997})}\BibitemShut {NoStop}%
\bibitem [{\citenamefont {Schecter}\ \emph {et~al.}(2016)\citenamefont
  {Schecter}, \citenamefont {Gangardt},\ and\ \citenamefont
  {Kamenev}}]{kamenev2016}%
  \BibitemOpen
  \bibfield  {author} {\bibinfo {author} {\bibfnamefont {M.}~\bibnamefont
  {Schecter}}, \bibinfo {author} {\bibfnamefont {D.}~\bibnamefont {Gangardt}},
  \ and\ \bibinfo {author} {\bibfnamefont {A.}~\bibnamefont {Kamenev}},\
  }\href@noop {} {\bibfield  {journal} {\bibinfo  {journal} {New J. Phys.}\
  }\textbf {\bibinfo {volume} {18}},\ \bibinfo {pages} {065002} (\bibinfo
  {year} {2016})}\BibitemShut {NoStop}%
\bibitem [{\citenamefont {Volosniev}\ and\ \citenamefont
  {Hammer}(2017)}]{volosniev2017}%
  \BibitemOpen
  \bibfield  {author} {\bibinfo {author} {\bibfnamefont {A.~G.}\ \bibnamefont
  {Volosniev}}\ and\ \bibinfo {author} {\bibfnamefont {H.-W.}\ \bibnamefont
  {Hammer}},\ }\href {\doibase 10.1103/PhysRevA.96.031601} {\bibfield
  {journal} {\bibinfo  {journal} {Phys. Rev. A}\ }\textbf {\bibinfo {volume}
  {96}},\ \bibinfo {pages} {031601 (R)} (\bibinfo {year} {2017})}\BibitemShut
  {NoStop}%
\bibitem [{\citenamefont {Mistakidis}\ \emph {et~al.}(2019)\citenamefont
  {Mistakidis}, \citenamefont {Volosniev}, \citenamefont {Zinner},\ and\
  \citenamefont {Schmelcher}}]{mistakidis2018}%
  \BibitemOpen
  \bibfield  {author} {\bibinfo {author} {\bibfnamefont {S.~I.}\ \bibnamefont
  {Mistakidis}}, \bibinfo {author} {\bibfnamefont {A.~G.}\ \bibnamefont
  {Volosniev}}, \bibinfo {author} {\bibfnamefont {N.~T.}\ \bibnamefont
  {Zinner}}, \ and\ \bibinfo {author} {\bibfnamefont {P.}~\bibnamefont
  {Schmelcher}},\ }\href
  {https://journals.aps.org/pra/abstract/10.1103/PhysRevA.100.013619}
  {\bibfield  {journal} {\bibinfo  {journal} {Phys. Rev. A}\ }\textbf {\bibinfo
  {volume} {100}},\ \bibinfo {pages} {013619} (\bibinfo {year}
  {2019})}\BibitemShut {NoStop}%
\bibitem [{\citenamefont {Dehkharghani}\ \emph {et~al.}(2018)\citenamefont
  {Dehkharghani}, \citenamefont {Volosniev},\ and\ \citenamefont
  {Zinner}}]{dehkharghani2018}%
  \BibitemOpen
  \bibfield  {author} {\bibinfo {author} {\bibfnamefont {A.~S.}\ \bibnamefont
  {Dehkharghani}}, \bibinfo {author} {\bibfnamefont {A.~G.}\ \bibnamefont
  {Volosniev}}, \ and\ \bibinfo {author} {\bibfnamefont {N.~T.}\ \bibnamefont
  {Zinner}},\ }\href {\doibase 10.1103/PhysRevLett.121.080405} {\bibfield
  {journal} {\bibinfo  {journal} {Phys. Rev. Lett.}\ }\textbf {\bibinfo
  {volume} {121}},\ \bibinfo {pages} {080405} (\bibinfo {year}
  {2018})}\BibitemShut {NoStop}%
\bibitem [{\citenamefont {Pastukhov}(2019)}]{pastukhov2018}%
  \BibitemOpen
  \bibfield  {author} {\bibinfo {author} {\bibfnamefont {V.}~\bibnamefont
  {Pastukhov}},\ }\href {https://doi.org/10.1016/j.physleta.2019.05.018}
  {\bibfield  {journal} {\bibinfo  {journal} {Phys. Lett. A}\ }\textbf
  {\bibinfo {volume} {383}},\ \bibinfo {pages} {2610} (\bibinfo {year}
  {2019})}\BibitemShut {NoStop}%
\bibitem [{\citenamefont {Panochko}\ and\ \citenamefont
  {Pastukhov}(2019)}]{pastukhov2019}%
  \BibitemOpen
  \bibfield  {author} {\bibinfo {author} {\bibfnamefont {G.}~\bibnamefont
  {Panochko}}\ and\ \bibinfo {author} {\bibfnamefont {V.}~\bibnamefont
  {Pastukhov}},\ }\href@noop {} {\bibfield  {journal} {\bibinfo  {journal}
  {arXiv:{\bf 1903.05953}}\ } (\bibinfo {year} {2019})}\BibitemShut {NoStop}%
\bibitem [{\citenamefont {Sacha}\ and\ \citenamefont
  {Timmermans}(2006)}]{sacha2006}%
  \BibitemOpen
  \bibfield  {author} {\bibinfo {author} {\bibfnamefont {K.}~\bibnamefont
  {Sacha}}\ and\ \bibinfo {author} {\bibfnamefont {E.}~\bibnamefont
  {Timmermans}},\ }\href {\doibase 10.1103/PhysRevA.73.063604} {\bibfield
  {journal} {\bibinfo  {journal} {Phys. Rev. A}\ }\textbf {\bibinfo {volume}
  {73}},\ \bibinfo {pages} {063604} (\bibinfo {year} {2006})}\BibitemShut
  {NoStop}%
\bibitem [{\citenamefont {Kain}\ and\ \citenamefont {Ling}(2016)}]{kain2016}%
  \BibitemOpen
  \bibfield  {author} {\bibinfo {author} {\bibfnamefont {B.}~\bibnamefont
  {Kain}}\ and\ \bibinfo {author} {\bibfnamefont {H.~Y.}\ \bibnamefont
  {Ling}},\ }\href {\doibase 10.1103/PhysRevA.94.013621} {\bibfield  {journal}
  {\bibinfo  {journal} {Phys. Rev. A}\ }\textbf {\bibinfo {volume} {94}},\
  \bibinfo {pages} {013621} (\bibinfo {year} {2016})}\BibitemShut {NoStop}%
\bibitem [{\citenamefont {Parisi}\ and\ \citenamefont
  {Giorgini}(2017)}]{parisi2017}%
  \BibitemOpen
  \bibfield  {author} {\bibinfo {author} {\bibfnamefont {L.}~\bibnamefont
  {Parisi}}\ and\ \bibinfo {author} {\bibfnamefont {S.}~\bibnamefont
  {Giorgini}},\ }\href {\doibase 10.1103/PhysRevA.95.023619} {\bibfield
  {journal} {\bibinfo  {journal} {Phys. Rev. A}\ }\textbf {\bibinfo {volume}
  {95}},\ \bibinfo {pages} {023619} (\bibinfo {year} {2017})}\BibitemShut
  {NoStop}%
\bibitem [{\citenamefont {Grusdt}\ \emph {et~al.}(2017)\citenamefont {Grusdt},
  \citenamefont {Astrakharchik},\ and\ \citenamefont {Demler}}]{grusdt2017}%
  \BibitemOpen
  \bibfield  {author} {\bibinfo {author} {\bibfnamefont {F.}~\bibnamefont
  {Grusdt}}, \bibinfo {author} {\bibfnamefont {G.~E.}\ \bibnamefont
  {Astrakharchik}}, \ and\ \bibinfo {author} {\bibfnamefont {E.}~\bibnamefont
  {Demler}},\ }\href {http://stacks.iop.org/1367-2630/19/i=10/a=103035}
  {\bibfield  {journal} {\bibinfo  {journal} {New Journal of Physics}\ }\textbf
  {\bibinfo {volume} {19}},\ \bibinfo {pages} {103035} (\bibinfo {year}
  {2017})}\BibitemShut {NoStop}%
\bibitem [{\citenamefont {Pastukhov}(2017)}]{pastukhov2017}%
  \BibitemOpen
  \bibfield  {author} {\bibinfo {author} {\bibfnamefont {V.}~\bibnamefont
  {Pastukhov}},\ }\href {\doibase 10.1103/PhysRevA.96.043625} {\bibfield
  {journal} {\bibinfo  {journal} {Phys. Rev. A}\ }\textbf {\bibinfo {volume}
  {96}},\ \bibinfo {pages} {043625} (\bibinfo {year} {2017})}\BibitemShut
  {NoStop}%
\bibitem [{\citenamefont {Kain}\ and\ \citenamefont {Ling}(2018)}]{kain2018}%
  \BibitemOpen
  \bibfield  {author} {\bibinfo {author} {\bibfnamefont {B.}~\bibnamefont
  {Kain}}\ and\ \bibinfo {author} {\bibfnamefont {H.~Y.}\ \bibnamefont
  {Ling}},\ }\href {\doibase 10.1103/PhysRevA.98.033610} {\bibfield  {journal}
  {\bibinfo  {journal} {Phys. Rev. A}\ }\textbf {\bibinfo {volume} {98}},\
  \bibinfo {pages} {033610} (\bibinfo {year} {2018})}\BibitemShut {NoStop}%
\bibitem [{\citenamefont {Astrakharchik}\ and\ \citenamefont
  {Pitaevskii}(2004)}]{astrakharchik2004}%
  \BibitemOpen
  \bibfield  {author} {\bibinfo {author} {\bibfnamefont {G.~E.}\ \bibnamefont
  {Astrakharchik}}\ and\ \bibinfo {author} {\bibfnamefont {L.~P.}\ \bibnamefont
  {Pitaevskii}},\ }\href {\doibase 10.1103/PhysRevA.70.013608} {\bibfield
  {journal} {\bibinfo  {journal} {Phys. Rev. A}\ }\textbf {\bibinfo {volume}
  {70}},\ \bibinfo {pages} {013608} (\bibinfo {year} {2004})}\BibitemShut
  {NoStop}%
\bibitem [{\citenamefont {Sykes}\ \emph {et~al.}(2009)\citenamefont {Sykes},
  \citenamefont {Davis},\ and\ \citenamefont {Roberts}}]{sykes2009}%
  \BibitemOpen
  \bibfield  {author} {\bibinfo {author} {\bibfnamefont {A.~G.}\ \bibnamefont
  {Sykes}}, \bibinfo {author} {\bibfnamefont {M.~J.}\ \bibnamefont {Davis}}, \
  and\ \bibinfo {author} {\bibfnamefont {D.~C.}\ \bibnamefont {Roberts}},\
  }\href {\doibase 10.1103/PhysRevLett.103.085302} {\bibfield  {journal}
  {\bibinfo  {journal} {Phys. Rev. Lett.}\ }\textbf {\bibinfo {volume} {103}},\
  \bibinfo {pages} {085302} (\bibinfo {year} {2009})}\BibitemShut {NoStop}%
\bibitem [{\citenamefont {Cherny}\ \emph {et~al.}(2012)\citenamefont {Cherny},
  \citenamefont {Caux},\ and\ \citenamefont {Brand}}]{Cherny2012}%
  \BibitemOpen
  \bibfield  {author} {\bibinfo {author} {\bibfnamefont {A.~Y.}\ \bibnamefont
  {Cherny}}, \bibinfo {author} {\bibfnamefont {J.-S.}\ \bibnamefont {Caux}}, \
  and\ \bibinfo {author} {\bibfnamefont {J.}~\bibnamefont {Brand}},\ }\href
  {\doibase 10.1007/s11467-011-0211-2} {\bibfield  {journal} {\bibinfo
  {journal} {Frontiers of Physics}\ }\textbf {\bibinfo {volume} {7}},\ \bibinfo
  {pages} {54} (\bibinfo {year} {2012})}\BibitemShut {NoStop}%
\bibitem [{\citenamefont {Paul}\ \emph {et~al.}(2005)\citenamefont {Paul},
  \citenamefont {Richter},\ and\ \citenamefont {Schlagheck}}]{Schlagheck2005}%
  \BibitemOpen
  \bibfield  {author} {\bibinfo {author} {\bibfnamefont {T.}~\bibnamefont
  {Paul}}, \bibinfo {author} {\bibfnamefont {K.}~\bibnamefont {Richter}}, \
  and\ \bibinfo {author} {\bibfnamefont {P.}~\bibnamefont {Schlagheck}},\
  }\href {\doibase 10.1103/PhysRevLett.94.020404} {\bibfield  {journal}
  {\bibinfo  {journal} {Phys. Rev. Lett.}\ }\textbf {\bibinfo {volume} {94}},\
  \bibinfo {pages} {020404} (\bibinfo {year} {2005})}\BibitemShut {NoStop}%
\bibitem [{\citenamefont {Tsuzuki}(1971)}]{tsuzuki1971}%
  \BibitemOpen
  \bibfield  {author} {\bibinfo {author} {\bibfnamefont {T.}~\bibnamefont
  {Tsuzuki}},\ }\href {\doibase 10.1007/BF00628744} {\bibfield  {journal}
  {\bibinfo  {journal} {Journal of Low Temperature Physics}\ }\textbf {\bibinfo
  {volume} {4}},\ \bibinfo {pages} {441} (\bibinfo {year} {1971})}\BibitemShut
  {NoStop}%
\bibitem [{\citenamefont {Ishikawa}\ and\ \citenamefont
  {Takayama}(1980)}]{ishikawa1980}%
  \BibitemOpen
  \bibfield  {author} {\bibinfo {author} {\bibfnamefont {M.}~\bibnamefont
  {Ishikawa}}\ and\ \bibinfo {author} {\bibfnamefont {H.}~\bibnamefont
  {Takayama}},\ }\href {\doibase 10.1143/JPSJ.49.1242} {\bibfield  {journal}
  {\bibinfo  {journal} {Journal of the Physical Society of Japan}\ }\textbf
  {\bibinfo {volume} {49}},\ \bibinfo {pages} {1242} (\bibinfo {year}
  {1980})}\BibitemShut {NoStop}%
\end{thebibliography}%

\onecolumngrid
\appendix

\section{Boundary-Augmented Cost Function}\label{app:J}
Equation~(6) shows the boundary-augmented cost function which includes a term that increases the cost for link-potential solutions that extend beyond the support region $x\in[-x_0,x_0]$. This added terms is
\begin{equation}
  J_{\mathrm{boundary}} = \alpha \sum_i^N J_{\mathrm{boundary}}^i,
\end{equation}
where 
\begin{equation}\label{eq:Jboundaryi}
  J_{\mathrm{boundary}}^i = \int\limits_{|x|>x_0}dx\,|V_i(x; A_i,\mu_i,\sigma_i)|^2.
\end{equation}
Assuming the Gaussian potential form as in Eq.~(5), this evaluates to
\begin{equation}\label{eq:JboundaryIGaussian}
  J_{\mathrm{boundary}}^i = \frac{\sqrt{\pi}}{2}A_i^2\sigma_i\left[
    \erfc\left(\frac{x_0+\mu_i}{\sigma_i}\right) +
    \erfc\left(\frac{x_0-\mu_i}{\sigma_i}\right)
  \right],
\end{equation}
where $\erfc$ is the complementary error function.

\section{Impurity in a Bose gas}\label{app:impurity_bose_gas}

To model one impurity atom that moves through a one-dimensional environment made of $N$ cold bosonic atoms, we employ the following Hamiltonian
\begin{equation}
H=-\frac{\hbar^2}{2m}\sum_{i=1}^N\frac{\partial^2}{\partial x_i^2}-\frac{\hbar^2}{2M}\frac{\partial^2}{\partial y^2}+\lambda \sum_{i>j=1}^N\delta(x_i-x_j)+g \sum_{i=1}^N \delta(x_i-y),
\end{equation}
where $M$ is the mass of the impurity atom, and $m$ is the mass of a bosonic particle. The position of the impurity is $y$, bosons are at the coordinates $\{x_i\}$. 
We assume that the realistic boson-boson and boson-impurity interactions are well-described by the zero-range potentials of strengths $\lambda$ and $g$, respectively. 
The environment is large by assumption. To describe it, the periodic boundary conditions are used: The particles move in a ring of the circumference $L$, such that $0<x_i<L$ and $0<y<L$.
We are interested in the thermodynamic limit: $N, L\to \infty$ with a fixed value of the density $\rho=\frac{N}{L}$.

If the system is non-interacting ($\lambda=g=0$), the eigenstates are written as $e^{2\pi i \frac{n_1x_1+\ldots+n_N x_N+my}{L}}$, where $n_1,\ldots,n_N$ and $m$ are arbitrary integers. 
For non-vanishing interactions we use these functions to write an eigenfunction of the Hamiltonian as $\Psi=\sum_{\{n_j\},m} a_{\{n_j\},m}e^{2\pi i \frac{\sum n_jx_j+my}{L}}$. 
Because all interactions are pairwise, 
the total (angular) momentum of the system must be conserved, and we write it as ${\cal P}=\frac{2\pi\hbar}{L}\left(\sum_{j} n_j+m\right)$.
A conserved quantity (${\cal P}$) allows us to exclude one variable from the consideration. 
We write the function $\Psi$ as $\Psi=e^{i \frac{{\cal P} y}{\hbar}}\sum_{\{n_j\},m} a_{\{n_j\},m}e^{2\pi i \frac{\sum n_jz_j}{L}}\equiv e^{i \frac{{\cal P} y}{\hbar}} \psi(z_1,...,z_N)$ 
with $z_i=L\theta(y-x_i)+x_i-y$, where $\theta(x)$ is the Heaviside step function, i.e., 
$\theta(x>0) = 1$ and zero otherwise. The variables $z_i$ are defined such that $0\leq z_i \leq L$ and the impurity is placed at $z=0$ ($z=L$). Now if we insert this 
function into the Schr{\"o}dinger equation, $H\Psi=E\Psi$, we obtain the following equation for $\psi(0<z_i<L)$
\begin{equation}
-\frac{\hbar^2}{2m}\sum_i\frac{\partial^2 \psi}{\partial z_i^2}-\frac{\hbar^2}{2M}\left(\sum_{i}\frac{\partial }{\partial z_i}\right)^2\psi
+ i \frac{\hbar {\cal P}}{M}\sum_{i}\frac{\partial \psi}{\partial z_i}+\lambda \sum_{i>j}\delta(z_i-z_j)\psi=\left(E-\frac{{\cal P}^2}{2M}\right)\psi,
\end{equation}
which must be supplemented with the boundary conditions:
\begin{equation}
\psi(z_i=0)=\psi(z_i=L); \qquad \frac{\partial \psi}{\partial z_i}\bigg|^{z_i=0^+}_{z_i=L^-}= \frac{2 g \kappa}{\hbar^2} \psi(z_i=0),
\end{equation}
where $\kappa=mM/(m+M)$ is the reduced mass.

\begin{figure}
 \centerline{\includegraphics[scale=0.45]{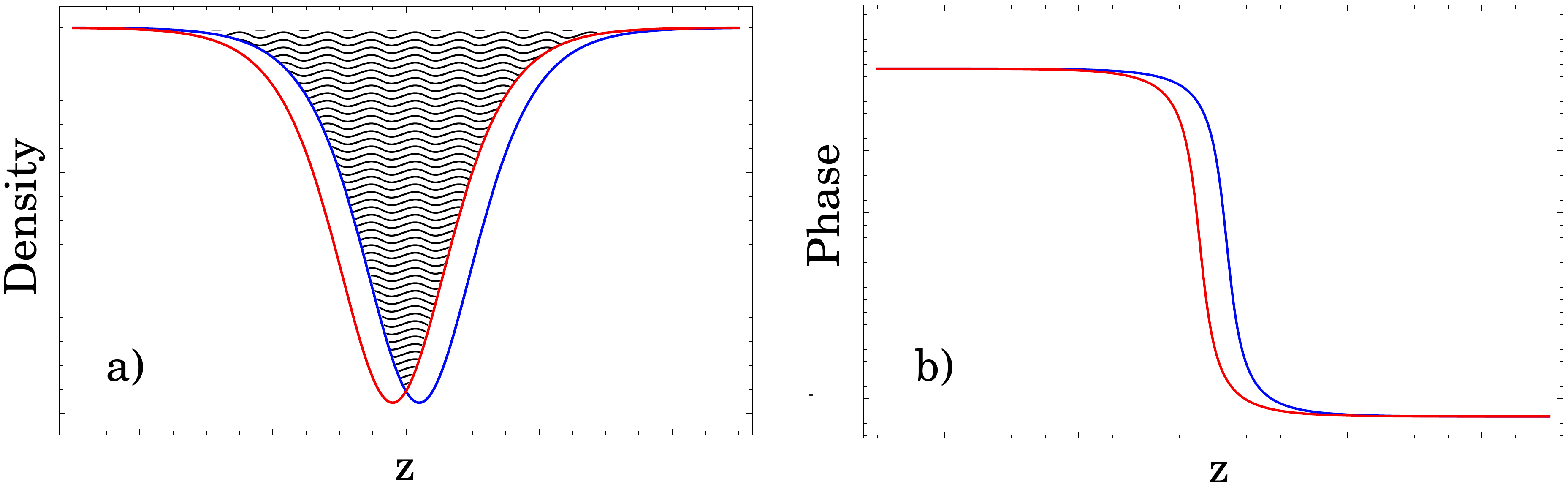}}
\caption{Panel {\bf a)}: The density, $|\Phi|^2 N$, of the Bose gas for $z_0=a$ (red curve), $z_0=-a$ (blue curve) 
($a>0$, the exact value of $a$ is not important for our discussion). 
 Note that the minimum of the density is at $-a$. 
The shaded area is a combination of the two solutions with the singularity at $z=0$. 
Panel {\bf b)}: The phase, $\phi$, of the Bose gas for the densities from {\bf a)}.}
 \label{fig:Fig1}
 \end{figure}

By assumption the bosons interact weakly, such that the ansatz $\psi=\prod_i \Phi(z_i)$ can be used to approximate the system. 
To minimize the expectation value of the Hamiltonian the function $\Phi(z)$ must satisfy the following non-linear Schr{\"o}dinger equation
\begin{equation}
-\frac{\hbar^2}{2\kappa}\frac{\partial^2\Phi}{\partial z^2}+i\frac{\hbar {\cal P}}{M}\frac{\partial \Phi}{\partial z}
-i \frac{\hbar^2 (N-1) A}{M} \frac{\partial\Phi}{\partial z} + \lambda (N-1)|\Phi|^2\Phi=\mu \Phi,
\end{equation}
where $A=-i\int \Phi(x)^*\frac{\partial}{\partial x}\Phi(x)\mathrm{d}x$ defines the momentum of a boson, and $\mu$ is the Lagrange multiplier. 
We rewrite this equation as
\begin{equation}
-\frac{\partial^2\Phi}{\partial z^2}+i v \frac{\partial \Phi}{\partial z} + \tilde \lambda (N-1)|\Phi|^2\Phi=\tilde\mu\Phi,
\label{eq:GPE_resc}
\end{equation}
where $\tilde\mu=\frac{2 \kappa \mu}{\hbar^2}$, $\tilde \lambda =\frac{2 \kappa \lambda}{\hbar^2}$, and 
$v= \frac{2 \kappa P}{M \hbar}$ with $P={\cal P}-\hbar A(N-1)$. $P$ defines the momentum of the impurity in the thermodynamic limit;
note that $A$ is determined by $P$, and there is a unique value of ${\cal P}$ for a given $P$.
The boundary conditions for Eq.~(\ref{eq:GPE_resc}) read
\begin{equation}
\Phi(z=0)=\Phi(z=L); \qquad \frac{\partial \Phi}{\partial z}\bigg|^{z=0^+}_{z=L^-}= \tilde g \Phi(0),
\end{equation}
where $\tilde g= \frac{2\kappa g}{\hbar^2}$. The non-linear equation~(\ref{eq:GPE_resc}) 
has an analytic steady solution~\cite{hakim1997}, which determines the properties 
of the dressed impurity in our problem. Let us first consider the non-interacting impurity $g=0$. 
In this case the solution for $v>0$ is~\cite{tsuzuki1971, ishikawa1980}
\begin{equation}
\Phi=\sqrt{\frac{\tilde \mu}{\tilde \lambda (N-1)}}\left(1-\beta \mathrm{sech}^2\left[\sqrt{\frac{\tilde\mu\beta}{2}}(z+z_0)\right]\right)^{\frac{1}{2}}e^{i\phi(z)},
\label{eq:Phi_c0}
\end{equation}
\begin{equation}
\phi(z)=-\pi\theta(z+z_d)+\mathrm{arctan}\left(\frac{\sqrt{\frac{2 v^2}{\tilde \mu}\beta}}{\mathrm{exp}\left[\sqrt{2\tilde \mu\beta}(z+z_0)\right]-2\beta+1}\right),
\label{eq:phi_c0}
\end{equation}
where  $\beta=1- v^2/(2\tilde \mu)$, $z_0$ is some parameter that determines the origin and $z_d$ is the point where $\mathrm{arctan}$ reaches $\pi/2$.
 It is worthwhile noting that the solution for $v<0$ is $\Phi^*$. 
The solution from Eqs.~(\ref{eq:Phi_c0}) and~(\ref{eq:phi_c0}) is plotted in Fig.~\ref{fig:Fig1}; 
for simplicity it is plotted in the interval $-L/2<z<L/2$, the region $0<z<L$ easily follows.

%%%%%%%%%%%%%%%%%%%%%%%%%%%%%%%%%%%%%%%%%%%%%%%%%%%%%%%%%%%%%%%%%%%%%%%%%%%%%%%%%%%%%%%%%%%%%%%%%%%%%%%%%%%%%%%%%%%%%%%%
 %%%%%%%%%%%%%%%%%%%%%%%%%%%%%%%%%%%%%%%%%%%%%%%%%%%%%%%%%%%%%%%%%%%%%%%%%%%%%%%%%%%%%%%%%%%%%%%%%%%%%%%%%%%%%%%%%%%%%%%

To describe an interacting impurity, we combine two moving solitons with $\pm z_0$, which creates a singularity at $z=0$~\cite{hakim1997,kamenev2016}.
Therefore, a dressed impurity in our model is a topological defect with a dissipationless propagation. 
We write the corresponding `wave function' as
\begin{equation}
\Phi=\sqrt{\frac{\tilde \mu}{\tilde \lambda (N-1)}}\left(1-\beta \mathrm{sech}^2\left[\sqrt{\frac{\tilde\mu\beta}{2}}(z\pm z_0)\right]\right)^{\frac{1}{2}}e^{i\phi(z)},
\end{equation}
with 
\begin{equation}
\phi(z)=\delta \phi \theta(-z)+\mathrm{arctan}\left(\frac{\sqrt{\frac{2 v^2}{\tilde \mu}\beta}}{\mathrm{exp}\left[\sqrt{2\tilde \mu\beta}(z\pm z_0)\right]-2\beta+1}\right),
\label{eq:phase_soliton}
\end{equation}
where $z_0>0$ is discussed below, the parameter $\delta \phi$ is not important for the further derivations, it reassures that the phase is a continuous function;
the plus sign in $\pm$ corresponds to $z>0$ and the minus sign to $z<0$. This function is illustrated in Fig.~\ref{fig:Fig1}. 
The density has a non-analytic derivative at $z=0$. The phase is a continuous function at $z=0$ (its derivative is also continuous).
Note that the wave function is not periodic (see Eq.~(\ref{eq:phase_soliton})). This non-periodicity is not important 
for our discussion, because we are interested in the behavior of the bosons close to the impurity. It suggests that a grey soliton
must be formed upon a change of interaction parameters to take care of the phase slip.

The parameter $\tilde \mu$ is found from the normalization condition $\int \Phi^2=1$. For $N\to\infty$, we obtain
\begin{equation}
\tilde \mu=\gamma \rho^2 \frac{N-1}{N}\left(1-2\sqrt{2\beta_0} \frac{ (\mathrm{tanh}(d)-1)}{\sqrt{\gamma}N}\right),
\end{equation}
where $\gamma=\tilde \lambda /\rho$, $\rho=N/L$, $\beta_0=1-v^2/(2\gamma \rho^2)$, and $d=\sqrt{\frac{\gamma \beta_0}{2}}\rho z_0$.
The equation to determine $z_0$ is found by using the boundary conditions at $z=\{0,L\}$
\begin{equation}
\frac{\tilde g}{\rho \sqrt{2\gamma}}=\frac{\beta_0^{\frac{3}{2}}\tanh(d)}{-\beta_0+\cosh^2(d)}.
\label{eq:c_v}
\end{equation}
This equation is cubic (in $\tanh(d)$), hence, the solutions can be found in a closed form.
There are three solutions. However, only two will lead to the acceptable values of $z_0$. 
We will refer to these steady solutions as the `polaron' and the `polaron-soliton' pair, because in the limit $g\to 0$ the former 
corresponds to the ground state, and the latter to a gray soliton. 
The `polaron-soliton' pair is expected to be unstable (small perturbations lead to 
a decay of this steady solution~\cite{hakim1997}),
therefore, we do not consider it. The solutions merge for $z_m$ 
\begin{equation}
\tanh^2\left(\sqrt{\frac{\gamma \beta_0}{2}}\rho z_m\right)=\frac{\sqrt{1+\frac{4v^2}{\gamma \rho^2}}-(1+\frac{v^2}{\gamma \rho^2})}{2\beta_0},
\label{eq:z_m}
\end{equation}
which is derived by taking a derivative of Eq.~(\ref{eq:c_v}) with respect to $z_0$ and equating the resulting expression to zero -- this determines 
the maximum value of $g$ for which (for a fixed $\beta_0$) there is a steady solution.
Equations~(\ref{eq:c_v}) and (\ref{eq:z_m}) give the equation for the critical value of $v_c$:
\begin{equation}
\frac{\tilde g}{\rho\sqrt{\gamma}}=\frac{3-\sqrt{1+\frac{4v_c^2}{\gamma \rho^2}}}{-1+\sqrt{1+\frac{4v_c^2}{\gamma \rho^2}}}\sqrt{\sqrt{1+\frac{4v_c^2}{\gamma \rho}}-1-\frac{v_c^2}{\gamma\rho^2}}.
\end{equation}
For $v>v_c$ (see Fig. 4 of the main text) there are no steady solutions.

Now we can calculate the energy of the dressed impurity in the thermodynamic limit
\begin{equation}
{\cal E}\equiv \lim_{N\to\infty, \frac{N}{L}\to \rho}\left[E(c,P)-E(c=0,P=0)\right],
\end{equation}
where
\begin{equation}
E(c,P) = \frac{{\mathcal P}^2}{2M} + \mu N-\frac{\hbar^2 A^2 N(N-1)}{2M}-\lambda N(N-1)\int_{0}^{L/2}|\Phi|^4\mathrm{d}z.
\end{equation}
Using these expressions we derive 
\begin{equation}
{\cal E}=\frac{P^2}{2M}+\frac{\hbar^2\rho^2}{2\kappa}\frac{\sqrt{2  \gamma \beta}}{3}\left[4 b + (-4b+\beta\mathrm{sech}^2(d))\tanh(d)\right]+\frac{\hbar P}{M}\lim_{N\to\infty} A N,
\end{equation}
where $b=1+\frac{v^2}{4\tilde \lambda \rho}=1+\frac{\kappa P^2}{2M^2 \lambda \rho}$. This energy for $v\to 0$ can be written as 
\begin{equation}
{\cal E}\simeq \epsilon+ \frac{P^2}{2m_{\mathrm{\mathrm{eff}}}},
\label{eq:18}
\end{equation}
where $\epsilon$ is the effective energy of the dressed impurity, and $m_{\mathrm{eff}}$ is the effective mass.

The parameters $m_{\mathrm{eff}}$ and $\epsilon$ calculated using Eq.~(\ref{eq:18})  
agree well with the results in the literature~\cite{mistakidis2018}, supporting 
the use of the non-linear Schr{\"o}dinger equation for solving the problem. 
However, further work is required to understand other properties of a dressed impurity in this formalism.
First of all, it will be interesting to investigate the critical momentum,
which is supersonic in the employed model for $g\to 0$ and for not heavy impurities.
Indeed, the model we solve is equivalent to a heavy impurity moving in a gas of bosons with mass $\kappa$,  
which has a different speed of sound. 
Note that Fig.~4 of the main text reports on a heavy impurity ($M/m\gg 1$) for which this problem does not occur.
It will also be interesting to investigate the residue -- the overlap between the wave function that describes 
a state with $g=0$ and the wave function that describes an interacting state. In the present model, the impurity changes the order parameter
only locally, which means a non-zero residue (see~\cite{volosniev2017} for $P=0$), contradicting other 
studies on the topic~\cite{pastukhov2017,grusdt2017}. To understand this disagreement, one could calculate the overlap using an exactly solvable model, e.g., 
a heavy impenetrable impurity in a Bose gas (solvable by the Bethe ansatz). 

\end{document}